# Schedulability Analysis in Time-Sensitive Networking: A Systematic Literature Review

Zitong Wang, Feng Luo, Yunpeng Li, Haotian Gan, and Lei Zhu

*Abstract*—**Time-Sensitive Networking (TSN) is a set of standards that provide low-latency, high-reliability guarantees for the transmission of traffic in networks, and it is becoming an accepted solution for complex time-critical systems such as those in industrial automation and the automotive. In time-critical systems, it is essential to verify the timing predictability of the system, and the application of scheduling mechanisms in TSN can also bring about changes in system timing. Therefore, schedulability analysis techniques can be used to verify that the system is scheduled according to the scheduling mechanism and meets the timing requirements. In this paper, we provide a clear overview of the state-of-the-art works on the topic of schedulability analysis in TSN in an attempt to clarify the purpose of schedulability analysis, categorize the methods of schedulability analysis and compare their respective strengths and weaknesses, point out the scheduling mechanisms under analyzing and the corresponding traffic classes, clarify the network scenarios constructed during the evaluation and list the challenges and directions still needing to be worked on in schedulability analysis in TSN. To this end, we conducted a systematic literature review and finally identified 123 relevant research papers published in major conferences and journals in the past 15 years. Based on a comprehensive review of the relevant literature, we have identified several key findings and emphasized the future challenges in schedulability analysis for TSN.**

## I. INTRODUCTION

WITH the development of industrial automation, various hard real-time systems in the industry need to transmit large amounts of data in real time to monitor, control, and optimize various processes [124]. For example, on a production line, sensors and actuators need to communicate with the control system in real-time to ensure the accuracy and efficiency of the production process [125]. Various automated devices also require high-bandwidth, low-latency communications to work together. In addition, with the development of intelligent vehicles, automobiles will carry many sensors and communication systems to sense their surroundings or coordinate between vehicles [126]. High bandwidth, low latency, and highly reliable data transmission are critical for these functions in hard real-time systems. Therefore, as a set of protocol clusters that provide low-latency and high-reliability guarantees, Time-Sensitive Networking (TSN) can play a crucial role in these areas.

This work was supported in part by the Shanghai Pudong New Area Science and Technology Development Fund Industry-University-Research Special Project (Future Vehicle) under Grant PKX2022-W01. *(Corresponding author: Zitong Wang).*

Zitong Wang, Feng Luo, Yunpeng Li, Haotia Gan, and Lei Zhu are with the School of Automotive Studies, Tongji University, Shanghai, 201804 China (e-mail: 1911048@tongji.edu.cn).

TABLE I
OVERVIEW OF THE TOOLS PROVIDING BOUNDED LOW LATENCY IN TSN STANDARDS

| Number | Main Content | Status |
|---|---|---|
| 802.1Qav | Credit Based Shaper | Published |
| 802.1Qbu | Frame Preemption | Published |
| 802.1Qbv | Scheduled Traffic | Published |
| 802.1Qch | Cyclic Queuing and Forwarding | Published |
| 802.1Qcr | Asynchronous Traffic Shaping | Published |
| P802.1Qdq | Shaper Parameter Settings | Ongoing |
| P802.1DC | QoS Provision | Ongoing |
| P802.1DU | Cut-Through Forwarding | Ongoing |
| P802.1Qdv | Enhancements to CQF | Ongoing |

TSN is an Ethernet-based real-time communication technology that extends the Audio Video Bridging (AVB) protocol [127]. While AVB ensures real-time transmission of audio and video streams, it lacks considerations for low-latency and high-reliability transmission of control traffic in real-time critical systems [128]. TSN adds richer scheduling and reliability assurance mechanisms to AVB and can provide high-bandwidth, low-latency, and reliable transmissions for the information interaction needs of hard real-time systems [129, 130]. It provides a valuable tool set for network designers in four areas: time synchronization [131], bounded low latency, reliable transmission [132, 133], and resource management [134, 135]. Network designers can use these mechanisms in TSN to make the network meet the needs of various real-time system scenarios. Among the tools provided by TSN, various scheduling mechanisms can provide Quality of Service (QoS) guarantees for traffic transmission, as shown in Table I.

The Credit-Based Shaper (CBS) proposed in IEEE 802.1Qav [136] can provide bounded audio and video traffic latency through the shaping policy with varying credit values. IEEE 802.1Qbu [137], in conjunction with IEEE 802.3br [138], provides the network with a frame preemption mechanism to guarantee low-latency transmission of time-critical traffic. The Time-Aware Shaper (TAS) in IEEE 802.1Qbv [139] avoids conflicts between streams by assigning appropriate transmission time windows, guaranteeing transmission determinism. To make the End-to-End (E2E) Latency of traffic in the network comply with deterministic requirements and be easy to compute, IEEE 802.1Qch [140] proposes the Cyclic Queuing and Forwarding (CQF), which allows correlating the computation of the E2E latency with the number of hops experienced by the streams. IEEE 802.1Qcr [141] also proposes Asynchronous Traffic Shaping (ATS) to provide bounded latency for transmitting burst traffic in the network. In addition to this, as can be seen from Table I, many standards are being proposed. For example, P802.1Qdq proposes scheduler parameter configurations under burst traffic, and P802.1Qdv targets enhancements to the CQF mechanism. However, since they are drafts rather than official documents, they are beyond the scope of this paper.

In hard real-time systems, the E2E latency of time-critical traffic must meet their deadline, or serious consequences will arise [142]. Therefore, the network designer must verify the timing predictability of the traffic in the network during the design phase, i.e., verify that the transmission of traffic in the network satisfies the specified timing requirements [143]. Therefore, when TSN is applied in time-critical systems, analyzing the network timing under various scheduling mechanisms in TSN is essential. Schedulability analysis in the network provides reliable upper bounds on traffic transmission latency [144]. It can test the reasonableness and validity of configuration parameters in the network, guiding the choice of scheduling mechanisms and queue mapping methods [143].

There are some core challenges in performing schedulability analysis for scheduling mechanisms in TSN. Firstly, there is a wide variety of methods for schedulability analysis. However, there is no work to categorize and compare them in detail, which leads to network designers being unable to master the analysis techniques and understand the effectiveness of each method. Secondly, multiple scheduling mechanisms in TSN may be combined, and the scheduling mechanisms that have not been analyzed yet remain to be clarified. Finally, it remains to be clarified under what network scenarios these scheduling mechanisms are analyzed and verified, which also affects the possible application scenarios of the various mechanisms. Although some current review work [126, 127] mentions schedulability analysis in TSN, this is only a part of their research, resulting in incomplete analyses and simplistic categorization of schedulability analyses. To the best of our knowledge, we are the first to conduct a Systematic Literature Review (SLR) on schedulability analysis in TSN.

This SLR identifies schedulable line analysis studies conducted in the field of TSN. The main objective of this SLR is to provide a detailed survey of recent advances in schedulability analysis research in TSN and to provide an entirely focused and well-organized literature classification scheme. This will help researchers and practitioners to identify and understand the existing state of the art of schedulability analysis in TSN and its applicability in various fields. Another contribution of this SLR is to identify the gaps in the current research on schedulability analysis in TSN and to clarify the direction of further research in the field.

We carried out a detailed analysis of current research and shortlisted 123 significant studies from an initial 789. We used a set of structured data extraction, analysis, and synthesis processes to drill down and collate these studies. Here is a summary of the key findings from our study:

- Most publications have been published by IEEE in conferences since 2012. The conference with the highest number of publications is the IEEE Conference on Emerging Technologies & Factory Automation (ETFA).
- The objective of all publications was to obtain tight bounds, with 37% of them evaluating scheduling mechanisms or achieving comparisons between mechanisms by obtaining tight boundaries.
- More than 80% of the publications did the schedulability analysis under CBS or TAS-related scheduling mechanisms, and most traffic classes analyzed were CDT and AVB traffic.
- The most used method for schedulability analysis in publications is Network Calculus (NC) (42% of the total), followed by Formal Timing Analysis (FTA), which does not provide a complete analytical framework.
- 50% of the publications were evaluated through Computational Analysis, followed by Analysis Combined with Simulation, where the most used simulation platform was OMNeT++.
- Supporting more mechanism and mechanism combinations, supporting more traffic classes, optimizing based on various schedulability analysis methods, and implementing validation methods combining simulations and comparisons are directions for future research.

The rest of the paper is organized as follows. Section II describes the methodology of this SLR, including a general overview of our methodology and a detailed review process. In Section III, we analyzed the publications' selection process results, and provided preliminary trend analysis and statistical examination of these collected publications. Subsequently, in Section IV, we present the data extraction strategy and its corresponding discovery. Section V will discuss the trends observed during our study and the challenges the community should strive to address. Section VI will outline the threats to the validity of this SLR. Section VII compares our findings with existing literature research, and finally, in Section VIII, we conclude this SLR.

## II. Methodology of This SLR

SLR is a systematic research methodology designed to comprehensively and objectively assess the existing literature in a specific research area. It systematically compiles, summarizes, and evaluates the relevant literature findings by defining a straightforward research question, defining a search strategy, selecting the literature, extracting data, and conducting a comprehensive analysis [145]. Referring to the classical SLR process proposed in the work in [146], we divide the adopted SLR process into three parts: Planning, Conducting, and Reporting, as shown in Fig. 1.

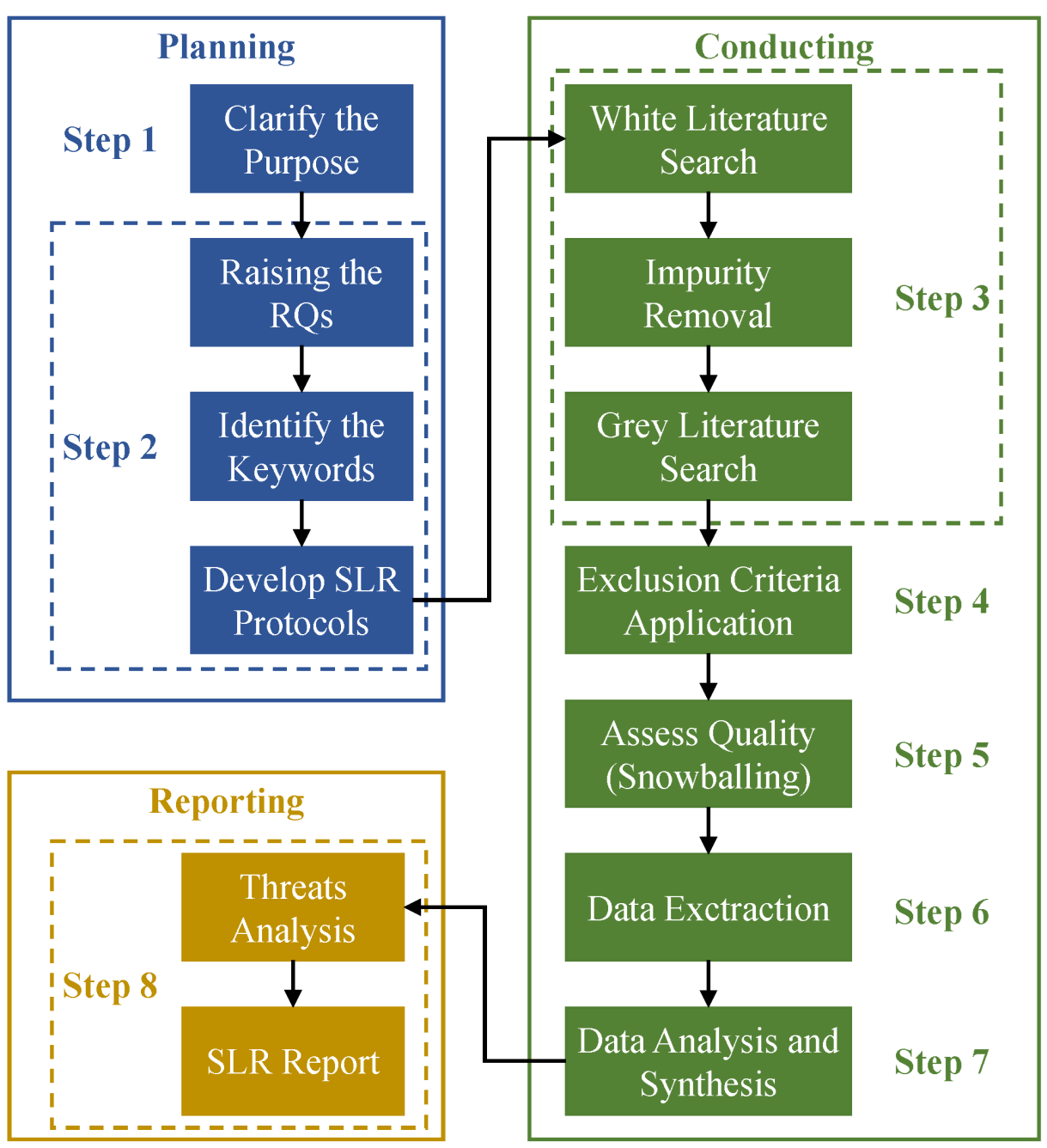

**Fig. 1.** Process of this SLR.

### *A. SLR Process*

As shown in Fig. 1, in the planning phase, we first identify the purpose for a review around schedulability analysis in TSN, define the core questions of this review, specify the keywords under study, and develop a review protocol that allows consistency in how the study members executed the review.

In the conducting phase, we draw on the grey search approach in [147] to implement the literature search. We first search for white literature and perform fusion and duplicate elimination on the white literature. A grey literature search then adds the literature. In the subsequent process, we screen the initial literature obtained from the search for exclusion based on the formulated conditions and evaluate them. In the evaluation process, we use the snowballing method proposed in [148] to add literature to prevent any missing literature from the search. The details of the grey literature search and snowballing are described in detail subsequently. At the end of this phase, we will extract, analyze, and synthesize the screened literature data.

In the reporting phase, we identified factors that could pose a threat to the validity of our study. We conducted a thorough threat verification and validation analysis [149]. The purpose of this paper is to document and describe in detail the research we conducted.

### *B. Research Purpose and Questions*

Given the possible challenges of schedulability analysis in TSN enumerated in Section I, we clarify the objectives of our review and formulate several research questions to highlight the research directions of the current literature to summarize and analyze them.

This SLR aims to classify the target scheduling mechanisms, methods, and scenarios, investigate research trends, identify strengths and weaknesses in the current state of the art, and emphasize the challenges and future directions for schedulability analysis research in TSN. To accomplish this purpose, we have identified the following research questions based on our understanding of the research area.

1) **RQ1:** What is the purpose of the schedulability analysis in these publications?

With this research question, we investigate the different goals pursued by schedulability analysis in TSN. In general, we investigated the goals of schedulability analyses at a high level to identify the requirements addressed in the literature (e.g., providing bounded delays, configuration recommendations, and building constraints.). We delve deeper into the specific features of schedulability analysis in TSN to enumerate the priorities that interest the current research.

2) **RQ2:** Which scheduling mechanisms and streams are targeted by schedulability analysis in TSN?

In the second research question, we investigate the classes of streams and scheduling mechanisms targeted by schedulability analysis in TSN to summarize which classes of streams in which scheduling mechanisms have been subjected to schedulability analysis at present and to classify them to reflect the current state of research and trends.

3) **RQ3:** What methods are used in schedulability analysis in TSN and how is the extensiveness of these methods?

In this research question, we investigated the different methods used for schedulability analysis in TSN, classified them based on their implementation, identified the object of study of each type of analysis method. In addition, we assessed the comprehensiveness of each method in considering the various types of blocking and the level of detail of their modeling. This will help us assess the effectiveness and extensiveness of different schedulability analysis methods and derive trends in these methods.

4) **RQ4:** How to evaluate the result of the schedulability analysis in TSN?

With this research question, we investigate the methods used in schedulability analyses in TSN to evaluate the results of the analyses. We summarize the methods used in the current study to assess the results of the analysis, which can be used as an essential indicator of the quality and reliability of the study. Moreover, we investigate the network scenarios chosen to assess results in schedulability analysis in TSN. We analytically summarize the application areas of the TSN used to evaluate the analysis results in the current study. This research question can also be divided into the following two sub-questions.

a) **RQ4.1:** What assessment methods are utilized for the schedulability analysis?

b) **RQ4.2:** What are the application domains in which TSN is being evaluated?

c) **RQ4.3:** What are the use case scenarios, topologies and traffic for evaluating the schedulability analysis.

TABLE II
SEARCH KEYWORDS USED IN THIS SLR

| Field | keywords | Search Field |
|---|---|---|
| Schedulability Analysis | Time Analysis, Timing Analysis, Delay Analysis, Latency Analysis, Schedulability Analysis, Maximum End-to-End, Worst Case End-to-End, Worst-Case End-to-End, Worst Case Response, Worst-Case Response, WCR, Worst-Case Delay, Worst Case Delay, WCD | All Field |
| Time-Sensitive Networking | Time Sensitive Network*, Time-Sensitive Network*, TSN, Audio Video Bridging, Audio/Video Bridging, AVB, Time-Aware Shaper, TAS, Credit-Based Shaper, Cyclic Queuing and Forwarding, Asynchronous Traffic Shaping, Urgency-Based Scheduler, IEEE 802.1Qav, IEEE 802.1Qb*, IEEE 802.1Qc* | Title-Abs-Key |

TABLE III
LITERATURE SOURCE USED IN THIS SLR

| Search Cat | Name | Type | URL |
|---|---|---|---|
| WL Search | Web of Science | Indexing system | www.webofknowledge.com |
| | IEEE Xplore Digital Library | Electronic database | www.ieeexplore.ieee.org |
| | ACM Digital Library | Electronic database | www.dl.acm.org |
| | SCOPUS | Indexing system | www.scopus.com |
| GL Search | ArXiv E-print Repository | Electronic database | www.arxiv.org |
| | Google Search Engine | Indexing system | www.google.com |

### *C. Review Protocols*

As shown in Fig. 1, this systematic literature review is conducted in two phases. First, in Step 4, we primarily review the abstracts and conduct a quick full-text scan to filter out irrelevant papers based on the defined criteria, resulting in preliminary screening results. Subsequently, in Step 5, we comprehensively review each publication and expand the literature set through the snowballing method. We extract relevant information to answer all the research questions (Step 6 in Fig. 1).

This study's co-authors follow the abovementioned literature screening and organization process. When there are disagreements in the analysis results of certain publications, we engage in informal discussions specifically addressing those publications until a consensus is reached [150, 151].

### *D. Search Keywords*

To quickly and easily search all the literature of interest, we first need to mark the keywords that need to be used for the search [152]. Our research focuses on schedulability analysis and TSN. Therefore, we need to find suitable keywords to cover the above two aspects, which include some precise terms and generalizations. The search keywords are shown in Table II.

Note that since CBS is proposed in AVB, we consider it also in the keywords for TSN. In addition, since schedulability analysis generally faces scheduling mechanisms in TSN, the keywords for TSN in this study only focus on adding the standard numbers related to scheduling mechanisms to simplify the search.

To make our search of the literature comprehensive while eliminating as much irrelevant literature as possible, we specify the search area for keywords related to TSN as "Title-Abs-Key" (represent the title, abstract, and keywords of the literature) and the search area for keywords related to schedulability analyses as "All Field." The full-area search for schedulability analyses is because there is literature that does not explicitly indicate the work associated with schedulability analyses in the title, abstract, or keywords. In the search process, the keywords in the two fields are operated as "and," and the keywords in each field are operated as "or." For example: Search_String = All_Field (SA.kw1 *or* SA.kw2 *or* …) *and* Title-Abs-Key (TSN.kw1 *or* TSN.kw2 *or* …). This will allow our search to cover the two fields' intersection fully. For databases or search systems that do not have xxx search patterns, such as the IEEE Xplore Digital Library and ACM Digital Library, we searched for titles, abstracts, and keywords separately. In addition, we restricted publications to a timeframe of the last 15 years (2009 to present). Then, we merged and de-duplicated them to obtain search results.

### *E. Literature Search*

Because SLR typically reviews only formally published literature and excludes the large amount of "Grey" Literature (GL) generated outside academic forums, SLR only provides a partial range of benefits [153]. The work in [154] points out that "White" Literature (WL) refers to sources where expertise and export control are well known. In contrast, grey literature refers mainly to the second tier of literature with moderate channel control and credibility. The work in [155] considers published journal papers, conference proceedings, and books as WL, preprints, e-prints, technical reports, lectures, data sets, Audio-Video (AV) media, and blogs classified as GL. In summary, we divide the literature search into two categories: white literature search and grey literature search. The sources of the two types of searches are shown in Table III.

To identify relevant research, at the time of the WL search, we performed an automated search in four of the largest and most complete databases in computer-related fields: Web of Science, the IEEE Xplore Digital Library, the ACM Digital Library, and Scopus as shown in Table III. These databases and systems are easy to access and can simplify the documentation steps as they support export functionality. In addition, they are widely recognized as one of the practical tools for conducting systematic literature reviews [156].

We mainly targeted the Google Search Engine and the ArXiv E-print Repository for the GL search, as shown in Table III. We used the exact search string for automated searches and a combination of manual searches with knowledge of the target research area. ArXiv is an open-access electronic pre-print (e-print) repository that aims to provide scientific researchers with free access to academic papers and research results [157].

*F. Selection Criteria*

After conducting the search based on the provided keywords, we proceeded to eliminate duplicate publications. However, even though these publications matched the keywords used in the search, some were found to be irrelevant to the topic of our study. Therefore, considering the potential inclusion of WL and GL, we have formulated detailed exclusion criteria to ensure a rigorous and consistent selection of publications by all members involved in the study, resulting in a more reliable set of literature. The Exclusion Criteria for WL (ECWL) and Exclusion Criteria for GL (ECGL) in this SLR are as follows.

- **ECWL1:** Papers in languages other than English were excluded because English is commonly used in scientific peer reviews worldwide.
- **ECWL2:** Secondary and tertiary studies such as surveys, conference reviews, and SLRs were excluded because they are usually inductive rather than proposing new techniques or programs.
- **ECWL3:** Standards were excluded due to their time-consuming nature and limited accessibility.
- **ECWL4:** Papers that are Work-in-Progress were excluded because they are often immature and may yield only some final results.
- **ECWL5:** Papers that could not be accessed in full text were excluded, as this affected our assessment of the quality of their research.
- **ECWL6:** Dissertations using results from conference or journal papers were excluded as this was considered duplicate research.
- **ECWL7:** Papers that did not perform the schedulability analyses were excluded. Some papers did not perform schedulability analyses but cited and validated analyses from other papers. Some papers only formally formulate for the E2E traffic latency without analyzing its worst-case scenario or using software to perform timing model checks. These papers were not considered in this SLR.
- **ECWL8:** Papers that did not perform schedulability analyses for traffic under the TSN scheduling mechanism were excluded. Some papers only optimized the methodology for schedulability analysis. Some papers only did schedulability analysis for traffic under the first-in-first-out mechanism (which is not a specific scheduling mechanism in TSN). These papers were not considered.

The Exclusion Criteria for GL (ECGL) are as follows.

- **ECGL1:** The web pages or preprint papers that did not perform the schedulability analyses were excluded.
- **ECGL2:** Videos, webinars, data sets, and books are excluded due to their time-consuming nature.
- **ECGL3:** The web pages or preprint papers that only summarize the current work were not considered.

TABLE IV
NUMBER VARIATION IN THE PUBLICATION SELECTION

| Step | Work Content | Count |
|---|---|---|
| 1 | After merge and impurity removal | 786 |
| 2 | After pre-screening (**ECWL1-5** and **ECGL1-2**) | 670 |
| 3 | After reviewing the titles or abstracts | 533 |
| 4 | After skimming or scanning the papers | 215 |
| 5 | After go through the full text | 139 |
| 6 | After final informal discussion | 115 |

After the above WL and GL search, we obtained 1166 publications. We eliminated 380 duplicates, resulting in 786 publications. Since it is relatively easy to determine whether a publication complies with **ECWL1-5** and **ECGL1-2**, we have divided the literature exclusion step into two steps: the exclusion of publications that comply with **ECWL1-5** and **ECGL1-2** by scanning the publications and the review of the remaining publications for compliance with **ECWL6-8** and **ECGL3** through an incremental approach.

The application of **ECWL1-5** and **ECGL1-2** resulted in 670 publications. Among them, 27 publications were excluded because they were not in English, and 37 publications were excluded because they were secondary or tertiary studies. Then, 26 standards and 7 Work-in-Progress papers were excluded, and 10 publications were excluded because the full text was unavailable. In addition, one web page was excluded because it only summarized existing work.

Next, we reviewed these 670 publications for compliance with **ECWL6-8** and **ECGL3** through an incremental approach. We eliminated papers that met the requirements of **ECWL6-8** and **ECGL3** by reviewing the title and abstract of each paper to determine whether it met the requirements of this SLR. After this step, the number of papers was reduced to 533. Then, we are identifying the presence of schedulability analyses for TSN scheduling mechanisms in publications by skimming or scanning the full paper. This step resulted in a smaller number of publications, to 215. Further, we went through the full text. However, in this process the authors disagreed on some of the publications. In order to prevent omissions, we considered all possibilities to obtain 139 potentially qualified publications. After informal discussions in response to these publications, we ended up with 115 publications, as shown in Table IV.

*G. Snowballing*

To minimize possible bias in the coverage of our study, we refer to [148] for snowballing. We searched the reference sections of the publications included in the 115 publications by backward snowballing. We tracked down publications that cited any of the publications included in the 115 publications by forward snowballing [157].

After the snowballing operation described above, we obtained eight more relevant publications. We show the citation relationships between them and the relevant publications, as shown in Fig. 2. In this Figure, each circle represents a publication, where the numbers represent its number (see Appendix for correspondence between numbers and citations).

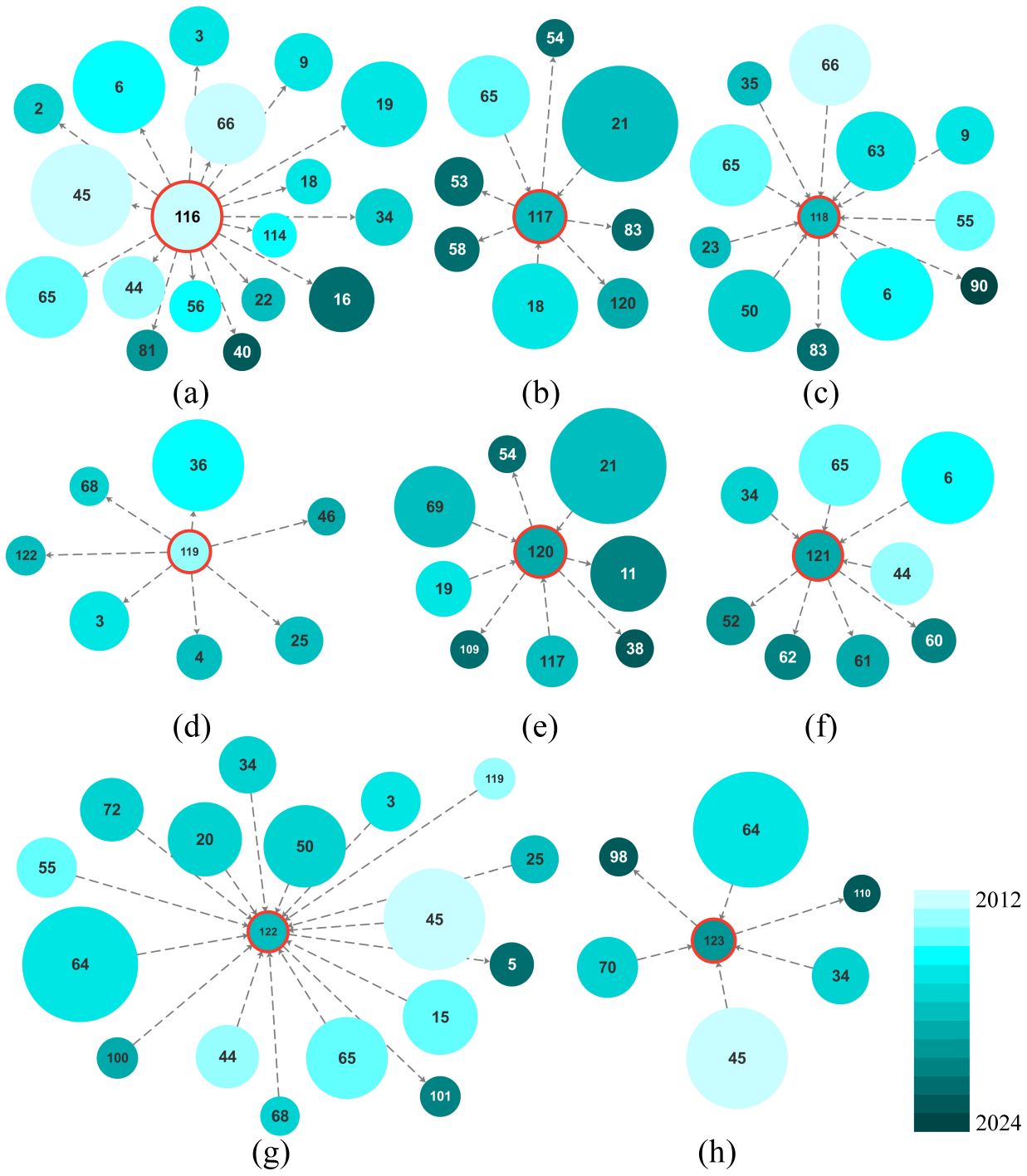

**Fig. 2.** Citation relationships of the publications in the snowballing process.

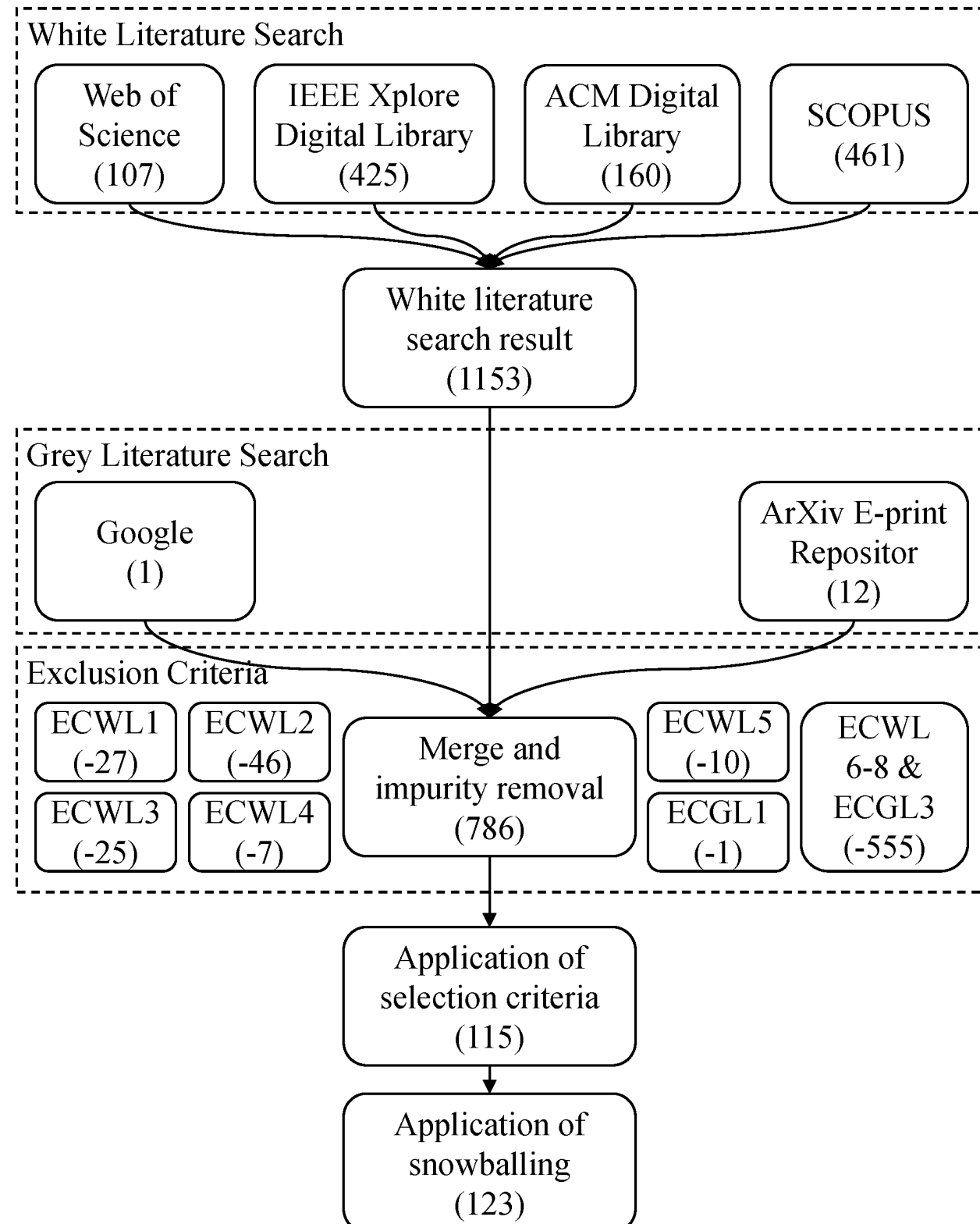

**Fig. 3.** Literature search and selection process in this SLR.

The size of the circle indicates the number of times the publication has been cited: the more citations, the larger the circle. The darker color of the circle indicates a later year of publication. The arrows indicate citation relationships between publications, with the publication at the end of the arrow citing the publication at the beginning of the arrow. For example, in Figure 2(a), it is shown that many other publications cite the publication numbered 116. In the end, we got the number of publications as 123. The number variation of publications throughout the search and selection process is shown in Fig. 3.

## III. Preliminary Analysis of the Publications

After the inclusion/exclusion screening and snowballing operations described above, we obtained 123 publications. In order to understand the general information, trends, and statistical characteristics of these publications, we conducted a preliminary analysis of them.

Fig. 4 shows the word cloud of journal or conference name abbreviations for all publications selected. The more times a conference appears, the more extensively its abbreviation is displayed in the word cloud. As can be seen from the figure, the conference with the highest relevance to this study is the IEEE Conference on Emerging Technologies & Factory Automation (ETFA). ETFA focuses on the latest developments and new technologies in the field of industrial and factory automation. The conference covers various topics, including information technology in automation, real-time and networked embedded systems, vehicular embedded systems, and industrial control. In addition, as can be seen in Fig. 4, the sources of publications include other conferences related to real-time communications, such as International Conference on Real-Time Networks and Systems (RTNS) and IEEE International Workshop on Factory Communication Systems (WFCS), as well as some journals that focus on electronic communications, such as IEEE Access and Real-Time Systems (RTS).

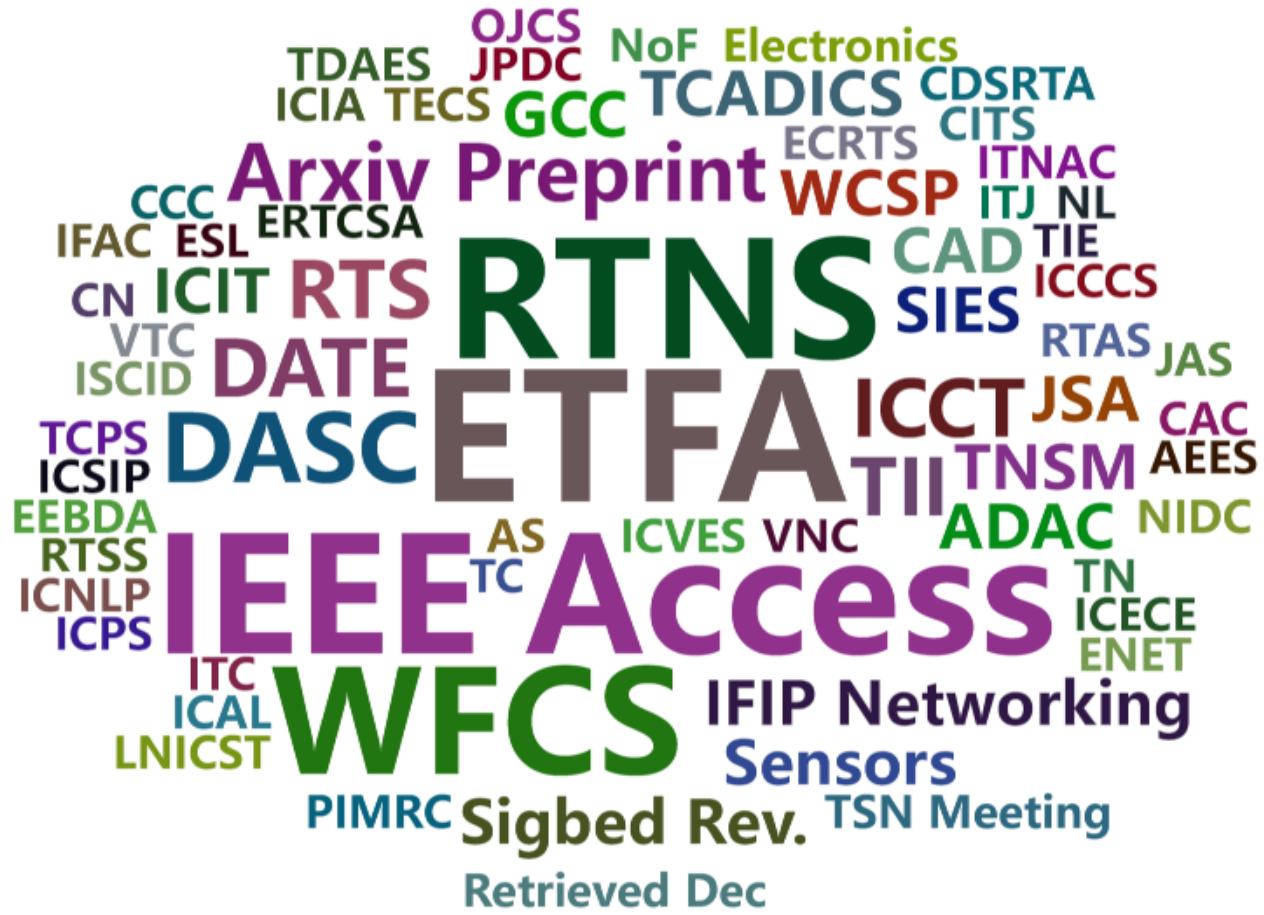

**Fig. 4.** Word cloud of conference or journal brief titles for all publications.

We have likewise counted the trend of the considered publications from year to year, as shown in Fig. 5. As seen in Fig. 5, even though we searched almost 15 years of publications, the earliest were only published in 2012. We can also see that although there is a decrease in some years (2013 and 2019), the number of publications analyzed for schedulability in TSN shows an overall upward trend as the number of years increases. The number of publications reached its highest in 2018 (19 publications) and has maintained high in recent years, demonstrating the value of research in this area.

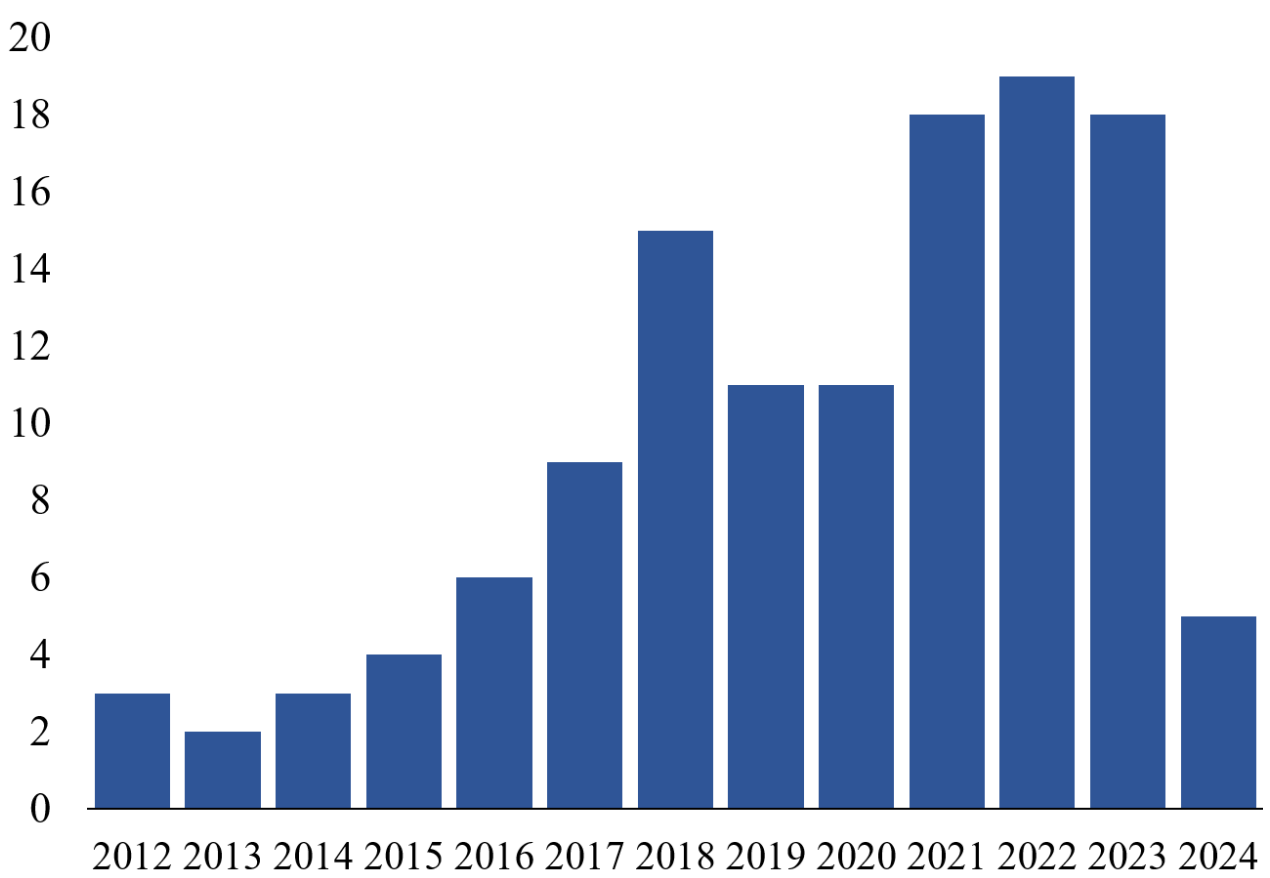

**Fig. 5.** Number of publications per year in the last 15 years.

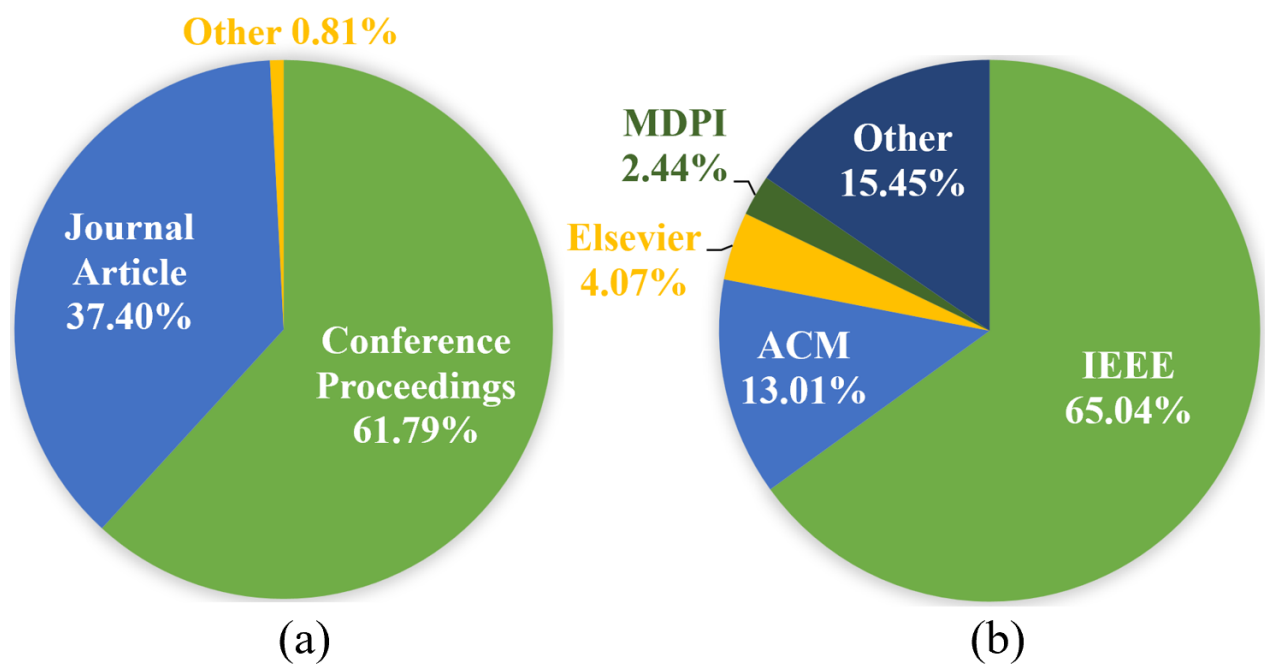

**Fig. 6.** Distribution of (a) types and (b) publishers of the publications

**Fig. 7.** Word cloud of author names for all publications.

In Fig. 6, we have categorized statistics on the types of publications and publishers. As shown in Fig. 6(a), the types of publications are mainly classified into journal articles and conference proceedings, of which 61.8% are published in academic conferences. This shows that most researchers want to shorten their publication cycle in case the real-time nature of their publications receives an impact.

From Fig. 6(b), IEEE published 65% of its publications because it has organized many high-quality conferences related to real-time communication, which indirectly contributes to the development of research on schedulability analysis in TSN. ACM publishes 13% of the publications, almost the same as other publishers' publications.

In addition, we have a comprehensive count of the authors of the publications, as shown in Fig. 7. Again, the more extensive the range of authors' names in Fig. 7, the more frequently they appear in the publications studied. In this figure, well-known scholars in TSN research, such as Luxi Zhao, Rolf Ernst, Paul Pop, and Ahlem Mifdaoui, have contributed to the field of schedulability analysis of TSN. This shows that these well-known scholars also focus on the schedulability analysis in TSN and regard it as a more state-of-the-art research direction. Fig. 7 also demonstrates that many current scholars are contributing to the analysis of schedulability in TSN.

## IV. Taxonomy of Schedulability Analysis in TSN

In order to answer the research questions posed in the above sections, we need to interpret all the publications in detail. From the research question, we propose a classification method as shown in Fig. 8; by classifying the information from the extracted publications, we can summarize to find the answer to the research question.

By analyzing the research questions mentioned above, we classify the research on schedulability analysis in TSN into five directions from a higher level: Analysis Objectives, Analysis Targets, Analysis Methodologies, Analysis Extensiveness, and Evaluation Scenarios. The Analysis Objectives are used to answer RQ1, the Analysis Targets is used to answer RQ2, the Analysis Methodologies and the Analysis Extensiveness are used to answer RQ3, and the Evaluation Scenarios is used to answer RQ4.

### *A. Analysis Objectives*

This categorization outlines the motivations and goals of schedulability analysis in the publications we studied. This section lists five possible purposes, including Tight Boundaries, Mechanism Evaluation, and Constraint Construction, Configuration Viability, as shown in Fig. 8.

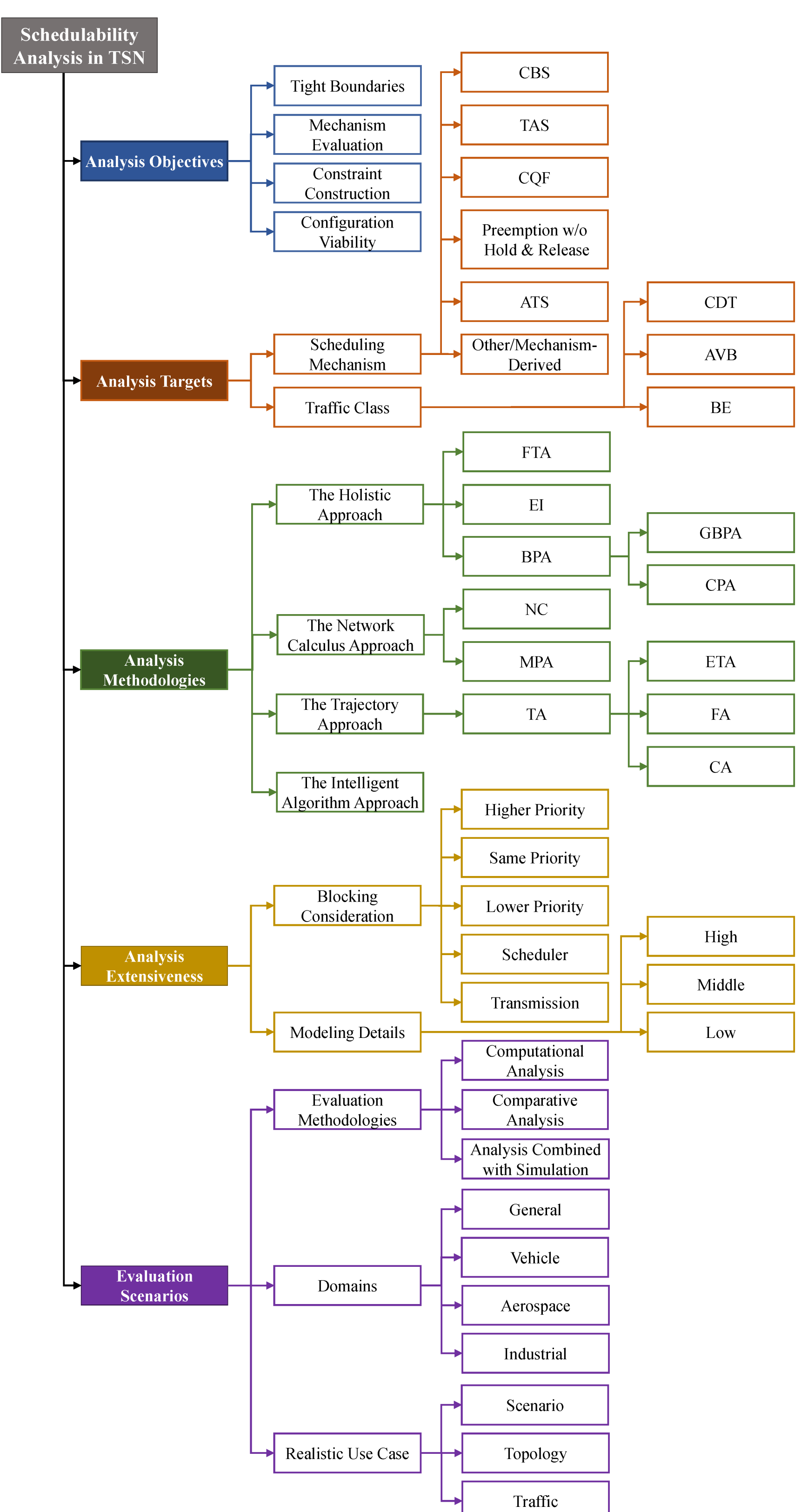

**Fig. 8.** Taxonomy of schedulability analysis of TSN

*B. Analysis Targets*

In order to explore the objects of schedulability analysis implementations in detail, we divide the Analysis Target into two main parts: scheduling mechanisms and traffic classes.

Among them, scheduling mechanisms are categorized according to the mechanisms currently supported by TSN. Note that the categorization focuses on whether the preemption supports the Hold & Release mechanism. At the same time, we have also considered other mechanisms, including those obtained through optimization based on existing mechanisms.

For traffic classes, we generalize the possible traffic classes in current communication networks: Control Data Traffic (CDT), AVB streams, and Best-Effort (BE) streams.

### *C. Analysis Methodologies*

Analysis methodology refers to the worst-case delay analysis methodology implemented by the publications. Among the analysis methodology, we consider the existing methodologies for schedulability analysis, referenced in [158], and classify them into four categories: the Holistic Approach, the Network Calculate approach, the Trajectory Approach, and the Intelligent Algorithmic Approach.

The Holistic Approach is analyzed for each forwarding device (switch) in the network, considering the worst-case scenario on each device through which the flow passes. The method outputs the best and worst-case scenarios of the flow as it passes through the previous forwarding device and uses them as inputs for the next forwarding device, thus obtaining the worst corresponding delay upper bound for the flow in the network. Therefore, the method is generally analyzed only for the worst-case scenario of the flow at one forwarding device. In order to analyze the publications in more detail, we classify the Holistic Approach into three categories by the difference in implementation methods: Formal Timing Analysis (FTA), Eligible Intervals (EI), and Busy Period Analysis (BPA), where BPA is subcategorized into General Busy Period Analysis (GBPA) and Compositional Performance Analysis (CPA).

The Network Calculus Approach considers a network element characterized by a service curve and an arrival curve for all flows accessing the device; it is possible to compute the maximum delay of any flow, the maximum size of the waiting queue, and the service curve of the flow. When the arrival and service curves are deterministic, then the results obtained by the method are deterministic. This SLR refers to the method based on arrival and service curves as the Network Calculus Approach. We categorize the Network Calculus Approach into Network Calculus and Modular Performance Analysis (MPA)

Unlike the Holistic Approach, the Trajectory Approach is based on analyzing the worst-case scenario experienced by the frame on its trajectory, not at any visited node. The method identifies the preceding frames and the busy periods that ultimately affect their latency one at a time by moving backward through the sequence of nodes visited by the analyzed frames. We categorize the Trajectory Approach into Extended Trajectory Approach (ETA), Forward End-to-End Delay Analysis (FA), and Composable Analysis (CA).

The Intelligent Algorithm Approach is quite different from the others in that it uses machine learning or deep learning algorithms to make predictions given traffic and network scenarios and thus analyze the schedulability of the traffic in the network. In the following, we describe each of these analysis methods.

### *D. Analysis Extensiveness*

Analysis Extensiveness is the level to which the analysis can be extended, indicating the generalization of schedulability analysis. For the Analysis Extensiveness of publications, we categorize the consideration from Blocking Consideration and Modeling Details. Blocking Consideration refers to the blocking that a publication considers when performing schedulability analysis, and we categorize it as Higher Priority Blocking, Same Priority Blocking, Lower Priority Blocking, Scheduler Blocking, and Transmission Blocking. We use the Modeling Details to denote the level of detail of the analysis and categorize it into High, Middle, and Low.

### *E. Evaluation Scenarios*

Evaluation Methodology refers to how a publication evaluates the results of its analyses. Evaluation Methodology is divided into three categories: Analysis of Computational Results, Compare with Other Analysis, and Analysis Combined with Simulation.

To further show the scenarios used in evaluating publication analysis results, we explore their evaluation domains. We categorized the domain into General, Vehicle, Aerospace, and Industrial to reflect the domain in which the scenarios used for the evaluation are located.

In addition, to summarize the scenarios constructed by the publications when evaluating the analysis, we explored their realistic use cases. We summarize the realistic use cases at three levels: scenario, topology and traffic.

## IV. SLR of Schedulability Analysis in TSN

In this section, we reflect on the results of our SLR by answering the RQs presented in the above sections.

### *A. What is the purpose of these schedulability analysis?*

We reviewed the objectives to be analyzed in the schedulability analysis in TSN. Schedulability analysis involves a variety of objectives, including obtaining bounded latency and backlogs for transmitting certain classes of traffic through the network, evaluating and comparing scheduling mechanisms, and acting as a constraint in the solution process, among others. In the following, we describe each of these objectives.

#### *1) Tight Boundaries*

In time-critical systems (hard real-time systems), each task must meet its deadline requirements, or irreparable damage may result. Therefore, the task-related behavior of the system must be predicted during its design to ensure its proper operation. The network tasks correspond to the transmission of streams, and the E2E latency of the streams also needs to meet specific latency requirements. Therefore, the most fundamental and shared purpose of schedulability analysis is to analyze the E2E latency of frames transmitted in the network and to derive an upper bound on the E2E latency of their transmission in the worst case. In some work, the E2E latency in the worst case is also called the Worst-Case Response Time (WCRT) [73, 75, 76] or the Worst-Case E2E Delay (WCD) [43, 47, 50]. A tight WCRT can provide network designers with a credible design basis, which can, in turn, improve network reliability and reduce resource wastage. Therefore, some works have tried to provide a tight WCRT for traffic in various network scenarios. For example, Cao *et al.* [33] proposed an alternative to the busy period approach that can more easily obtain a tight WCRT for AVB traffic in the presence of low-priority or high-priority interference. Subsequently, they continued their work to make the method capable of dealing with more complex situations and obtaining a tighter WCRT [75, 76]. Benammar *et al.* [77] extended the FA approach to support analyses under the CBS mechanism and concluded that the method could obtain a tighter WCRT through comparisons. Moreover, the backlog is another attribute that needs to be considered during the implementation of time-critical systems. Since the size of the buffer is related to the difficulty and cost of the implementation, it is a non-negligible factor in real applications. Therefore, providing an upper bound for backlog is also one of the aims of all the publications.

*2) Mechanism Evaluation*

In network communication, different application scenarios may impose different requirements on traffic transmission, which also requires further investigation of the effectiveness of different scheduling mechanisms in different network application scenarios. E2E latency, as one of the critical network performance metrics, is often used to evaluate the effectiveness of scheduling mechanisms. Therefore, the WCRT obtained in schedulability analysis can also be used as a metric to evaluate the scheduling mechanism. Once the upper bound of E2E latency and backlog are obtained, they can be used as metrics for mechanism comparison and evaluation to assess whether a scheduling mechanism is suitable for a specific scenario. AS an example, Docquier *et al.* [17] and Seliem *et al.* [52] performed schedulability analysis of scheduling mechanisms in TSN, respectively, as a way of exploring their potential application in industrial domains such as Substation Automation Systems (SAS) and Industrial Internet of Things (IIoT). In addition, some studies have also evaluated the performance of scheduling mechanisms for other application scenarios, such as Avionics Full Duplex Switched Ethernet (AFDX) [37, 38] and in-vehicle networks [2, 16]. WCRT from the schedulability analysis can be used not only to verify whether a mechanism can meet the application requirements in different scenarios but also to compare scheduling mechanisms, such as TAS and ATS [43], CBS and Strict Priority (SP) [79], and TAS and CQF [39].

*3) Constraint Construction*

Much attention has been paid to solving configuration parameters and queue assignments for some of the current scheduling mechanisms in TSN, particularly TAS [126]. Constraints usually need to be constructed during the solving process to make the results of the solution usable. Among these constraints, the WCRT derived from schedulability analysis can be used to construct constraints related to E2E latency so that the resulting parameters can satisfy the latency requirements of the network [1, 35, 42, 114]. For example, the queue assignment problem for real-time streams over TSN was formalized as a Satisfiability Modulo Theory (SMT) specification in [1], which proposed a WCRT analysis for constructing the Real-Time Constraint in solving the queue assignment problem. In addition to solving for configuration parameters and queue assignments, some of the work also guides the design during traffic transmission path (routing) planning by analyzing the schedulability of certain traffic classes in TSN to ensure that critical traffic can meet the timing requirements [21, 102]. To determine the routing of AVB flows and minimize their worst-case end-to-end delay, Laursen *et al.* [21] combined schedulability analysis with a heuristics algorithm based on the Greedy Random Adaptive Search Process (GRASP).

*4) Configuration Viability*

In order to realize real-time communication in time-critical systems, devices in the network need to be set up with appropriate scheduling mechanisms, and the scheduling mechanisms need to have appropriate configuration parameters. Through schedulability analysis, it is possible to check whether the current parameters can meet the latency requirements of all time-critical traffic in the network without simulation or prototyping. Thus, schedulability analysis can be used to support and validate the parameter configuration in TSN, such as the configuration of reserved bandwidth [87, 100] and slope [18, 82] in CBS. For example, Hasan *et al.* [18] predicted the overall traffic conditions for different flows by analyzing the behavior of the credits under different timing conditions and finally analyzes the impact of the slope configuration on the traffic delivery by evaluating it with examples. Kong *et al.* [82] combined the schedulability analysis by proposing a deadline-aware slope allocation algorithm to compute the bandwidth required on each link to satisfy the deadlines of all flows. Moreover, to efficiently achieve practical configurations, some work also uses methods such as machine learning to explore the configuration space using CBS [5, 81, 84, 119].

*5) Insights from RQ1*

As shown in Table V, providing tight boundaries is the primary purpose of the schedulability analysis. This is also because deadline is an essential metric in hard real-time systems and is the core of schedulability analysis. More attention should be paid to verifying and implementing schedulability analysis in TSN. A fully functional software timing modeling platform can significantly reduce the difficulty of network development and provide design guarantees for network developers. In addition, most of the work provides bounds mainly on the latency upper bounds of traffic transmission. In contrast, only a tiny portion of the work focuses on the backlog, which is an integral part of the implementation process and affects the network design from the perspective of hardware resources.

TABLE V
SUMMARY OF THE PUBLICATIONS ACCORDING TO THE ANALYSIS PURPOSE

| Analysis Objectives | Reference to the paper | Count |
|---|---|---|
| Mechanism Evaluation | [2, 3, 6, 7, 10, 12, 16, 17, 19, 20, 23-28, 30, 32, 34, 36-40, 43, 52-54, 58, 59, 62, 64, 65, 67, 74, 79, 89, 97, 107-113] | 45 |
| Constraint Construction | [1, 9, 11, 13, 21, 29, 35, 42, 57, 60, 70, 71, 102, 105, 106, 114, 117] | 17 |
| Configuration Viability | [4, 5, 8, 18, 19, 45, 66, 68, 80-82, 84, 87, 89-92, 95, 100, 103, 104, 108, 119] | 24 |

Our categorical analysis also shows that in addition to obtaining tighter delays, many studies have further used the WCRT derived from schedulability analyses for mechanism evaluation and comparison, constructing constraints in the solution process, and evaluating the feasibility of parameter configurations. This also shows that these fields are the most concerning issues in the research addressing the mechanism aspects of TSN scheduling. The application of schedulability analysis in these aspects can reduce the difficulty of evaluating the mechanism's practicality and parameter configuration's difficulty. At the same time, it can provide real-time guarantees for algorithmic solutions such as parameters or routing.

*B. Which streams and scheduling mechanisms are targeted by schedulability analysis in TSN?*

We reviewed the stream classes and scheduling mechanisms in the schedulability analysis in TSN. In the following, we describe each of these stream classes and mechanisms.

*1) Stream Classes*

TSN supports multiple classes of traffic and their combinations. The traffic classes of streams studied in this literature review primarily include CDT, AVB, and BE.

*a) CDT*

CDT is usually used in hard real-time applications that require low latency and low jitter. The data frames for these flows are usually transmitted cyclically. For example, sensor data must be sent periodically to a computing unit within a cutoff time. Control data frames are usually assigned the highest priority to ensure deterministic delivery behavior throughout the network. Similarly, this type of traffic is also the most important and needs to be scheduled, so it is also called ST (Scheduled Traffic) in some work. CDT is usually considered deterministic because it can be controlled at the sending node. Since the CDT is the most critical traffic in the network, its latency is often used to confirm whether the timing requirements are met under specific scheduling mechanisms, such as BLS [36-38], TAS [56, 57], and ATS [20, 58, 60].

*b) AVB*

AVB traffic is typically real-time audio and video streams that are non-critical but have stringent bandwidth requirements, and their data frames are intermittent and large in size. AVB traffic is usually bursty. This traffic class requires consistent allocation of network resources to ensure quality of service requirements. Examples include video streams for detecting products in an industrial network and video streams transmitted in an automotive entertainment system. AVB streams are usually shaped by the CBS mechanism [33, 74, 75]. Many publications also study the schedulability of AVB streams under the combined mechanisms [2, 98, 120]. For example, Wang et al. [2] proposed a method to obtain the WCRT of AVB traffic under CBS and TAS. Moreover, they proved it through experiments in Vector CANoe.

TABLE VI
SUMMARY OF THE PUBLICATIONS ACCORDING TO THE ANALYZED TRAFFIC CLASS

| Traffic class | Reference to the paper | Count |
|---|---|---|
| CDT | [1, 7, 8, 10-13, 16, 17, 20, 23-25, 29, 30, 32, 35-72] | 54 |
| AVB | [2, 3, 15, 18, 21, 22, 30-34, 40, 41, 43, 45, 47, 49, 52, 54, 65, 74-80, 82, 83, 85-94, 98-100, 102, 103, 106, 108, 109, 111, 113, 117, 120-123] | 54 |
| BE | [1, 4, 41-44, 52, 73, 80, 96, 98] | 11 |

*c) BE*

BE traffic streams do not require timing guarantees and are usually mapped to low priority. Since BE streams do not require guaranteed latency, few publications are focusing on their schedulability analysis. Smirnov *et al.* [96] gave a schedulability analysis on BE traffic in automotive TSN networks using the TAS mechanism. Furthermore, since CBS introduces an additional delay to the BE stream, the work in [73] analyzed the BE stream's WCRT to explore that effect.

We categorized the initial publications according to their analyzed traffic class, as shown in Table VI. Note that some publications do not specify the traffic class they analyzed, such as [114], [97], and [115]. We do not categorize them.

*2) Scheduling Mechanisms*

TSN provides a comprehensive suite of scheduling and shaping mechanisms tailored to different traffic requirements. For instance, periodic and time-sensitive traffic often necessitates the use of synchronous scheduling mechanisms such as TAS, CQF, and CBS, which are specifically designed to maintain temporal consistency. In contrast, non-periodic or sporadic time-sensitive traffic can effectively utilize Frame Preemption or ATS mechanisms, which allow for greater flexibility in handling non-periodic traffic. Moreover, the combination of different mechanisms can be strategically employed to further refine the shaping of traffic. Additionally, research has also delved into the various derivatives of these shaping mechanisms to explore their potential benefits and applications.

*a) CBS*

CBS, introduced by IEEE 802.1Qav, is usually used to shape AVB streams. It controls the rate at which traffic is sent by assigning credits to specific traffic types to ensure that real-time traffic in the network is transmitted on demand. It should be noted that BLS has not been included in the standard, but since its basic principle is the same as CBS, it is classified as CBS in this article. Apart from BLS [36-38], which is commonly used for shaping CDT, CBS is often associated with AVB streams. CBS is longer established compared to other shaping mechanisms, making it well-established in the field of TSN. Therefore, the schedulability analysis of CBS is extensive. For example, Diemer *et al.* [93] used CPA to obtain the WCRT of streams under the CBS mechanism. Different from [93], Mohammadpour *et al.* [88] used NC approach to analyze the latency and backlog bounds under the CBS mechanism. Moreover, Li *et al.* [78] developed a worst-case delay analysis in the context of AVB switched Ethernet networks using ETA, which can provide tighter bounds than [88] and [93].

*b) TAS*

TAS is a scheduling mechanism introduced by IEEE 802.1Qbv. TAS uses time-based gates at the output ports of devices to shape the traffic. The network designer can guarantee the certainty of its transmission by designing a suitable GCL to send the traffic within a specific time window. As a time-triggered scheduling mechanism in TSN, many schedulability analyses exist for the TAS mechanisms [39, 48, 50]. In addition to the analyses of traffic under the TAS mechanisms only, there is much work that has investigated the WCRT in the case of integration of TAS and CBS [22, 41, 100]. Moreover, Some publications integrated the mechanisms of preemption based on CBS and TAS and analyzed their schedulability, such as [33], [48], and [49].

*c) CQF*

CQF can be regarded as an application of TAS, which is realized by alternating receiving and sending by the priority queues. This mechanism can provide bounded E2E latency with a simplified latency calculation method. Few current studies on schedulability analyses under the CQF mechanism exist. Instead of using the FTA approach [16, 20] to formalize the WCRT for traffic under the CQF mechanism (also called the Peristaltic Shaper), Thiele *et al.* [39] used CPA to characterize the timing of traffic under this mechanism. Luo *et al.* [107] analyzed the WCRT for various classes of traffic in CQF and preemption integration scenarios based on [39] , and their subsequent work [159] validates and analyzes the conclusions drawn in [39] to obtain parameter configuration suggestion for CQF and preemption integration scenarios.

*d) Preemption*

Frame preemption is a scheduling mechanism introduced by IEEE 802.1Qbu. Frame preemption allows express frame to preempt preemptable frame, which can provide a reduced latency transmission for scheduled, time-critical frames. Since preemption can change the frame transmission timing, the schedulability analysis for this mechanism is also necessary [44]. Most other timing analyses are based on this research, such as [107] and [98]. When Frame preemption is used in conjunction with TAS, Hold & Release can be introduced to reduce the length of guard band. Therefore, schedulability analysis for preemption and TAS usage can be categorized as with [52, 102, 122] and without [40, 47, 98, 120] Hold & Release.

TABLE VII
SUMMARY OF THE PUBLICATIONS ACCORDING TO THE ANALYZED SCHEDULING MECHANISMS

| **Mech.** | **Reference to the paper** | | **Cnt** |
|---|---|---|---|
| | **Individual** | **integrated** | |
| CBS | [3, 4, 15, 16, 18, 31-33, 36-38, 46, 49, 54, 55, 59, 62, 65-67, 72-95] | [2, 5, 21, 22, 34, 40, 41, 43, 45, 47, 52, 98-100, 102, 103, 106, 108, 109, 111, 117, 119-123] | 44/26 |
| TAS | [8, 10, 11, 13, 16, 17, 19, 20, 23-25, 29, 30, 32, 39, 48, 50, 51, 56, 57, 61, 63-65, 68, 69, 71, 96] | [2, 5, 7, 12, 21, 22, 27, 40, 41, 43, 45, 47, 52, 98-100, 102, 103, 106, 108, 109, 111, 114, 117, 119-123] | 28/29 |
| CQF | [16, 20, 39] | [107] | 3/1 |
| Preemption w/o H&R | [28, 44] | [5, 7, 21, 26, 34, 40, 47, 52, 98, 102, 103, 107, 119, 120, 122] | 2/15 |
| ATS | [6, 20, 58, 60, 88, 104, 110, 112] | [27, 43] | 8/2 |
| Other/ Derived | [1, 9, 14, 32, 35, 42, 53, 54, 70, 81, 95, 97, 101, 105, 113, 115, 116, 118] | [12, 26, 108, 114] | 18/4 |

*e) ATS*

ATS, defined by IEEE 802.1Qcr, offers a unique advantage in that it can effectively manage network traffic without relying on time synchronization, making it suitable for scenarios in which precise timing is not as critical or in which there are challenges in achieving global synchronization across the network. ATS is based on eligible time to shape traffic. There are few publications now focusing on the ATS mechanism. For example, Specht *et al.* [58] introduced a scheduling algorithm named Urgency-Based Scheduler (UBS), in which ATS was originally proposed, and presents a timing analysis on this mechanism using GBPA. Zhao *et al.* [43] evaluated the performance of ATS using NC and analyzed different mechanism integrations, such as ATS+CBS, TAS+ATS, and TAS+ATS+CBS.

*f) Other / Mechanism-Derived*

Other mechanisms in TSN, such as Strict Priority (SP) [84], Per-Stream Filtering and Policing (PSFP) [26, 108, 114] and Packet Replication and Elimination Redundant (FRER) [97], can impact the scheduling results in addition to the ones described above, which are, in turn, taken into account in the schedulability analysis. Some publications have also proposed optimized shaping mechanisms and conducted timing analyses thereof. For example, Ojewale *et al.* [115] considered the impact induced by multi-level preemption mechanisms. Additionally, to minimize the interruption of SR traffic on BE traffic by reducing the reserved bandwidth for SR traffic, Han *et al.* [108] introduced a traffic scheduling algorithm integrated with ingress shaping and used the NC approach to analyze the schedulability.

*3) Insights from RQ2*

Table VI and Table VII show that in the current research on schedulability analysis for scheduling mechanisms in TSN, AVB is the most analyzed traffic, and CBS is the most analyzed mechanism, followed by TAS. Only a little work has been done on schedulability analysis for other scheduling mechanisms, such as CQF and ATS.

Due to the complex nature of data traffic in TSN networks, a multifaceted approach is often required to leverage various mechanisms' strengths for optimal network performance. Researchers often look beyond investigating single scheduling and shaping mechanisms to explore how different mechanisms can be combined synergistically. As we can see from Table VII, there are a large number of publications focusing on the combination of mechanisms. Many studies combine TAS and CBS, such as [109] and [45]. For example, Ashjaei *et al.* [45] gave a schedulability analysis of the combination TAS, CBS and frame preemption (both with and without Hold & Release are considered). The combination of frame preemption with other mechanisms has also been a focus in schedulability analysis. As mentioned above, most timing analysis of frame preemption are based on the work in [40]. As an example, Luo *et al.* [107] provided a schedulability analysis using CPA under CQF with preemption. Regarding the theoretical research in [107], they then provided the simulation verification using OMNeT++ [159].

*C. What methods are used in schedulability analysis in TSN and what is the difference between them?*

In this SLR, we answer this research question by combining the Analysis Extensiveness with the Analysis Methodology in the classification of Fig. 8. In the following, we describe each of these approaches.

*1) Analysis Methodologies*

The methods used for schedulability analysis in TSN can be broadly classified as the holistic, NC, trajectory, and intelligent algorithmic approaches. Fig. 8 shows some more detailed branches under each.

*a) FTA*

FTA analyzes traffic timing by formalizing network and traffic-related parameters. This method is more straightforward than other methods. For example, Wang *et al.* [2] used the FTA to analyze Stream-Reservation (SR) type traffic under the combined use of TAS and CBS. They formalized the WCRT into three components: interference in the CBS mechanism, interference in the First-in-First-Out (FIFO) queue, and interference from the TAS mechanism, and described them through other detailed formalisms. Similarly, Zhao *et al.* [3] formalized the delay of this traffic class through the FTA, which is considered as the sum of the blocking caused by the CDT, the blocking caused by the guard band, and the blocking caused by the CBS. Moreover, Thangamuthu *et al.* [16] obtained the WCRT of CDT under BLS, TAS, and CQF by formalizing parameters such as non-CDT frame length, the residence time of a CDT frame inside a switch, and maximum time between two CDT slots. This analysis approach is elementary and does not consider complex scenarios, and the formalization needs to be more detailed. From the above, most of the focus of FTA is on classifying streams subjected to blocking and their formal representation and less on analyzing the timing behavior of streams.

*b) EI*

EI was first proposed to provide tight delay upper bounds for AVB streams (SR Class) under CBS in [33]. The method studies the so-called eligible intervals, defined by the method as the period during which frames are pending to be sent and the corresponding shaper is allowed to send them. The method only studies the interference within the available period to find the maximum latency experienced by any frame relative to the no-interference case. Cao *et al.* [33] stated that in an Ethernet AVB switch, interference to AVB traffic, whether higher priority or lower priority blocking, will delay at most one interfering frame compared to the no interference case. However, it also points out that EI-based analysis cannot analyze the case of simultaneous blocking of both higher and lower priority streams. To address this issue, they analyzed the more complex case where the interference combines higher and lower priority streams in [75]. However, the scenario analyzed in this work still needs to be improved: it analyzes the scenario consisting of a single-shaped higher priority stream and an unshaped lower priority stream. This scenario reduces the generalizability of this analysis, and reliable results can only be obtained in specific scenarios. Therefore, the further optimization of [76] to the publication [75] proved that more than the conclusions drawn from the analysis in [75] are needed to solve this problem. The conclusions drawn from the analysis in [75] still hold in the presence of multiple high-priority streams. In addition, Cao *et al.* [87], combined with the schedulability analysis of SR streams via EI, recommended bandwidth reservation. Moreover, the EI approach can also be used to obtain the WCRT of BE streams under CBS [73] or AVB streams under the combination of TAS and CBS [109].

*c) GBPA*

In this SLR categorization, we consider both GBPA and CPA to be BPA because they are both concerned with the blocking suffered by traffic during busy periods, leading to their worst-case latency. The busy period sis the maximum time interval during which a network resource is busy. Note that a resource is considered busy when there is a transmission in progress on the link or when the queue is not empty but cannot be transmitted due to the scheduling mechanism. Some publications also refer to this method as Response-Time Analysis (RTA) [35, 98]. GBPA can be used not only to analyze traffic under separate scheduling mechanisms, such as CBS [4, 74, 82] and ATS [58, 60], but also in cases where scheduling mechanisms are used in combination, such as CBS and TAS with [40, 98] or without [45] preemption. For example, to provide bounded latency for SR Class streams under CBS, Bordoloi *et al.* [74] used GBPA to analyze the streams of Class A and B to derive a formal expression for the worst-case response time. Subsequently, Kong *et al.* [82] builds on [74] (requiring only the necessary conditions for schedulability) by analyzing all SR Class streams using GBPA and using the results of the analysis to guide the design of traffic routing.

*d) CPA*

CPA is one of the main methods of formal timing analysis for deriving upper bounds on the worst-case end-to-end latency. It divides real-time embedded systems' timing models into task, component, and system timing [160]. In order to meet deadlines in real-time systems, the worst-case response time of a task needs to be obtained by a local analysis that considers the scheduling policies. In the local analysis, a parameter called level-i busy period is introduced to indicate the time interval during which the resource processes tasks or tasks with higher priority [161], which shows that CPA is an extension of BPA. Diemer *et al.* [93] transformed the AVB network model into a timing analysis model and used CPA to derive the worst-case timing characteristics and timing guarantees of AVB. Then, they utilized CPA to analyze the performance of distributed embedded systems and the worst-case timing of streams in AVB (under SP and CBS mechanisms) [54]. Building on this, Axer *et al.* [83] extended and generalized the ideas in [54] to provide tighter bounds under all conditions when using CBS. In addition to the analysis of CBS, some work has also analyzed other mechanisms using CPA, such as FIFO [101], BLS [36], TAS [39, 96], preemption [93], and CQF [39]. Moreover, CPA can also be used in mechanism integration cases in TSN [107].

*e) NC*

NC is a theory of systems used to figure out guaranteed upper bounds on latency by modeling flows and nodes as functions. It analyzes individual elements of a network, such as individual queues, complete nodes, or the entire network [162]. This approach generally uses curves to abstract the incoming data and schedule it for each system so that bounds can be derived for each system [163]. Queck *et al.* [86] analyzed the delay upper bounds for streams under the CBS mechanism using NC for the first time. It derived service curves for two CBS streams and investigates the increase in stream bursts after each hop. Subsequently, Azua *et al.* [85] used detailed arrival curves, additional services, and shaping curves based on the work in [86], making the results tighter. Then, there is a partial optimization study on that work [15, 89, 120, 122, 123]. For example, Zhao *et al.* [15] enhanced the previous work considering a third CBS class. Then, they obtained service curves in scenarios where TAS and CBS are combined and worst-case response times for CBS traffic in scenarios with and without preemption [120].

*f) MPA*

Real-Time Calculus (RTC) extends NC's basic concepts to the field of real-time embedded systems based on NC's theoretical framework [164]. MPA enhances RTC with a flexible component model; arrival curves can be constructed based on components' behavior without obtaining the service curves [80]. Reimann *et al.* [80] presented the MPA approach to analyzing AVB traffic in complex Electronic/Electrical Architectures (EEA) in automobiles. It compares the analytically derived latency upper bounds with simulations to demonstrate their reliability.

*g) ETA*

The Trajectory Approach (TA) obtains the worst-case response delay of a message transmission by analyzing it over its transmission path. Unlike the holistic approach, TA can take the effect of the previous hop into account when considering the worst-case response time of the traffic. Li *et al.* [78] proposed ETA to extend the definition of the busy period based on TA to improve the computation of worst-case end-to-end delay bounds of streams in CBS. However, some work has pointed out that the TA cannot obtain the worst-case latency when the cumulative load of the stream along its path exceeds 100% [165].

*h) FA*

Like TA, FA exploits the limitations imposed by the upstream nodes that will be considered when performing the theoretical analysis. It propagates the busy period analysis along the stream path. It has the advantage over TA by lifting the load restrictions. Benammar *et al.* [77] used the FA approach and extends the Strict Priority mechanism to support traffic analysis under the CBS mechanism. They compared this result with those of the CPA analysis to verify the effectiveness of the analysis.

*i) CA*

CA is an extension of FA. Unlike the FA approach, CA uses a compositional approach to formally express the maximum time for frames to complete transmission during the analysis process. Kong *et al.* [99] analyzed AVB traffic using the CA approach and derives an upper bound on the delay of traffic transmission. The work states that while the analysis using CA will improve the speed of the solution compared to existing work, it also introduces a certain amount of pessimism into the analysis.

*j) Intelligent Algorithm*

The intelligent algorithm approach refers to verifying the schedulability of the mechanisms through machine learning and deep learning. The main objective of this approach is to verify the feasibility of the TSN configuration. For example, Mai *et al.* [81] explored whether traditional schedulability analysis can be replaced by machine learning. The work employs the k-nearest neighbors (k-NN) approach and compares it to NC-based schedulability analysis regarding accuracy and efficiency. Meanwhile, they showed that machine learning effectively predicts the feasibility of actual TSN networks [95]. Then, they proposed using a predictive uncertainty metric to minimize the rate of false positives [84]. In addition, their work in [5] proposed a prediction model based on the graph neural network (GNN) for schedulability analysis, which will be improved in [119] in terms of training set construction and model structure.

TABLE VIII
SUMMARY OF THE PUBLICATIONS ACCORDING TO THE ANALYSIS METHODS

| Method | Reference to the paper | Count |
|---|---|---|
| FTA | [2, 3, 6-32] | 29 |
| EI | [33, 73, 75, 76, 87, 109] | 6 |
| GBPA | [1, 4, 35, 40, 42, 45, 58, 60, 74, 82, 98, 103, 105, 114, 118] | 15 |
| CPA | [36, 39, 44, 53, 54, 83, 93, 96, 101, 107, 111, 115] | 12 |
| NC | [34, 37, 38, 41, 43, 46-52, 55-57, 59, 61-72, 79, 85, 86, 88-92, 94, 97, 100, 102, 104, 106, 108, 110, 112, 113, 116, 117, 120-123] | 52 |
| MPA | [80] | 1 |
| ETA | [78] | 1 |
| FA | [77] | 1 |
| CA | [99] | 1 |
| Intel. Algo. | [5, 81, 84, 95, 119] | 5 |

*2) Analysis Extensiveness*

Schedulability analysis is usually done by formally modeling the network and considering various worst-case blocking to obtain the WCRT for the traffic. Its extensiveness not only impacts the practical effectiveness of the application but also determines the system's capacity to handle increasing network sizes and more complex traffic characteristics. Schedulability analysis primarily relies on the development of network and traffic models to account for potential traffic blocking within the network. Therefore, when assessing the extensiveness of the analysis, we consider it in terms of blocking considerations and modeling details.

*a) Blocking Consideration*

Since frames are transmitted based on strict priority regardless of the scheduling mechanism in TSN, the blocking a frame suffers is usually analyzed by priority in schedulability analysis. Therefore, in this SLR, we classify the blocking experienced by frames in schedulability analysis into higher-priority blocking (HPB), same-priority blocking (SPB), lower-priority blocking (LPB), and scheduler blocking (SCB). Frames also experience transmission delays (TD) while being sent. In an SP-based transmission process, higher priority frames can usually be sent before the frame of the analyzed stream starts or finishes (depending on whether preemption is supported) transmission. Because of the FIFO principle, same-priority traffic arriving before the arrival of the frame under analysis can produce blocking. Lower-priority frames usually cause blocking only for the duration of the longest frame transmission (related to the scheduling mechanism). Similarly, different schedulers may introduce different blocking for frames. The transmission delay is the time the analyzed frame has to experience during transmission. It is usually related to the transmission rate and the length of the frame. Although schedulability analysis should consider all types of blocking, publications typically assume a predetermined network scenario in their analyses. This assumption may limit the comprehensiveness of the analysis with respect to all potential blocking types, thereby influencing the applicability of schedulability analysis across different contexts.

TABLE IX
SUMMARY OF THE PUBLICATIONS LACKING OF CONSIDERATION OF SOME BLOCKING

| Blocking Consideration | Reference to the paper | Count |
|---|---|---|
| Without HPB | [3, 10, 12, 13, 16, 19-21, 23-29, 31-34] | 19 |
| Without SPB | [3, 4, 9, 11-13, 15, 16, 19-29, 31, 32] | 20 |
| Without LPB | [3, 4, 7, 9, 10, 12, 13, 15, 19-21, 23-29, 31, 73, 105] | 21 |
| Without SCB | [11, 19] | 2 |
| Without TD | [3, 16, 21] | 3 |

During the analysis, some publications assume that the traffic class being analyzed belongs to the lowest or the highest priority and, therefore, does not consider the LPB [4] or the HPB [34]. In the scenarios under consideration, the traffic class being analyzed does not account for higher or lower priority streams, thereby limiting the scalability of the analysis to more complex scenarios, such as those involving multiple priority streams within a traffic class. Some publications also provide a formal description of the delay experienced during a frame's transmission without formalizing the blocking caused by streams of other priorities. In the scenario they have constructed, the stream transmission delay is solely comprised of the individual components of the transmission process, without accounting for the potential interference caused by other traffic in the network that may block its transmission. For example, Nsaibi *et al.* [19] only formalized the delay experienced by traffic under different topological segments without further refining the various types of blocking that traffic may experience. Note that Cao *et al.* [33] proposed an EI approach to analyzing AVB traffic. They constructed scenarios that do not simultaneously incorporate both high-priority and low-priority streams. Therefore, the WCRT analysis in this publication did not consider the co-existence of HPB and LPB (they optimized for this in their subsequent work). For ease of representation, we have classified it as the category that does not consider HBP in this SLR. Our categorization of publications according to the type of blocking not considered is shown in Table IX.

TABLE X
SUMMARY OF THE PUBLICATIONS ACCORDING TO THE LEVEL OF MODELING DETAILS

| Level | Reference to the paper | Count |
|---|---|---|
| Middle | [1-14] | 14 |
| Low | [16, 17, 19-25, 27, 28, 31, 32] | 13 |

*b) Modeling Details*

For schedulability analysis, incorporating more detailed parameters during the modeling phase enhances the applicability of the analysis to a wider range of network scenarios. As modeling also affects the extensiveness of analysis, we rated publications according to the level of detail in the modeling. In schedulability analysis, a comprehensive model must encompass all relevant traffic attributes and key network mechanism parameters. These parameters include: the static characteristics of the traffic (e.g., load length or transmission time), the transmission characteristics (e.g., period and jitter), and the network configuration parameters (particularly those within the TSN mechanism that influence traffic transmission). Based on these parameters, we classify the level of modeling detail into three categories: high, medium, and low. A "high" level of modeling detail signifies that both traffic and network characteristics are thoroughly modeled. A "medium" rating indicates that one category is omitted or that some parameters were not fully considered. A "low" rating indicates that only one category was considered or that some parameters were significantly underrepresented. Note that we regard this rating as an assessment of the correctness of publications employing the IA methods, as these publications typically do not incorporate the parameters described above. The rating is performed by comparing the degree of correctness across these publications.

As we can see from Table X, most publications have a high level of modeling detail, which refers to their inclusion of all scheduling mechanism-related and stream attribute-related parameters. However, some publications did not model the static characteristics or transmission parameters of the relevant streams during the analysis, so their level of detail is rated as Middle. For example, Wang *et al.* [2] formally described the WCRT for AVB traffic, which considered parameters such as the Idle Slope of the CBS mechanism. However, they did not model the transmission characteristics of streams; in other words, the impact of the period and jitter of the streams is not considered in the analysis. Note that we treat this rating as an assessment of the correctness of publications that use the IA approach since they do not usually involve parameter modeling. In addition, some publications did not address the parameters related to scheduling mechanisms and stream properties in their modeling process but simply described the stream's delay formally. For example, Thangamuthu *et al.* [16] formalized the blocking experienced by messages in switches applying different TSN mechanisms but did not model the properties of the analyzed streams. Although some parameters of the scheduling mechanism are involved in the analysis, they remain at a low level of modeling detail. Our categorization of publications according to the level of modeling details is shown in Table X.

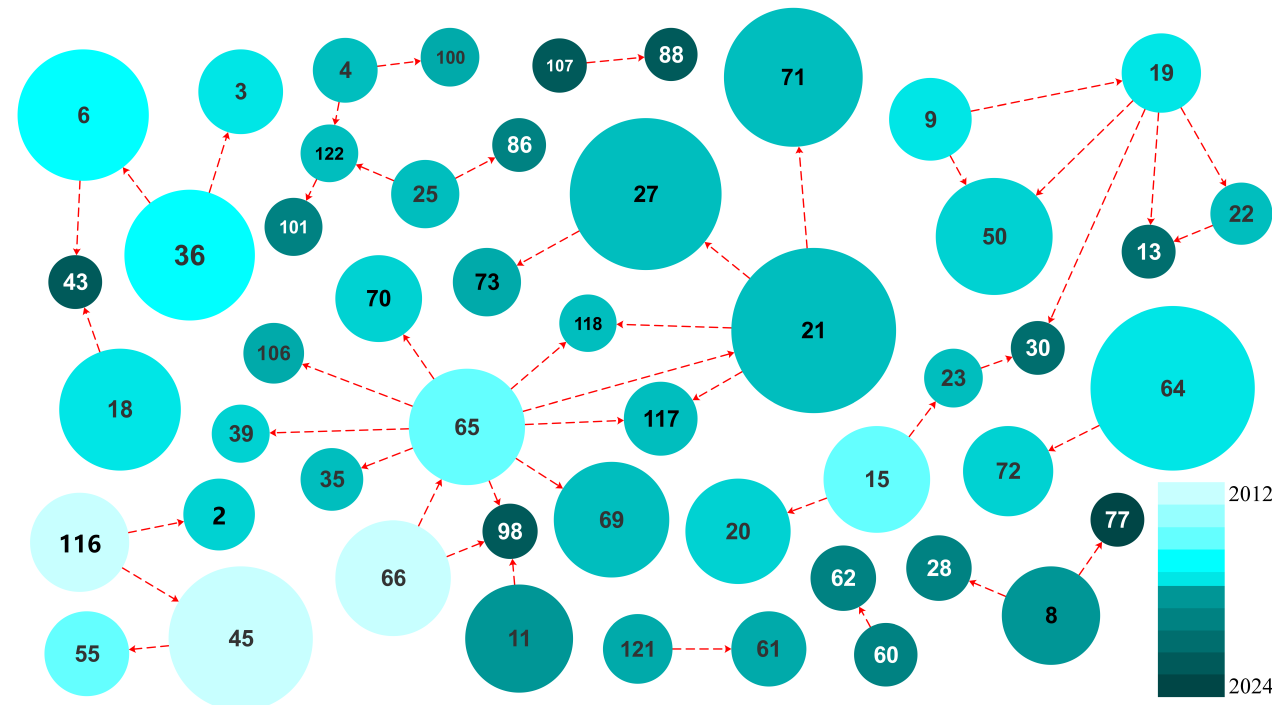

**Fig. 9.** Extended relationship between the publications.

*3) Insights from RQ3*

As Table VIII shows, NC is the most used approach for schedulability analysis in TSN, followed by busy-period approaches such as GBPA and CPA. The FTA, FA, and CA approaches have only one study for each, as they were proposed later. Only a few publications use intelligent algorithmic approaches to predict schedulability, but as this technology develops, more research may be done using this approach.

While many of the publications consider all possible blocking completely, some still need to express some blocking in a granular manner, which will affect the effectiveness of the schedulability analysis. In addition, some publications need to fully consider the scheduling mechanism's parameters and traffic attribute models, which may result in a lack of generality in their schedulability analyses.

Moreover, we further explored the relationship between the extensions of each publication, as shown in Fig. 9. Similar to Fig. 2, the color and size of the circles represent the year of publication and the number of citations, respectively. In this figure, the starting end of the arrow refers to the base publication, and the trailing end is the expanded publication. For example, as stated above, Azua *et al.* [85] used several curves based on the work in [86], making the analysis results tighter. Therefore, the circle numbered 65 is connected to the circle numbered 66 by an arrow pointing to the circle numbered 65. Combined with Table VIII, most of the expansion between publications involves expansion between the same analysis methods.

As can be seen in Fig. 9, many of the schedulability analyses through the NC approach are based on [85] and [86] (number 65 and 66) for the extensions. The publications related to the EI approach are extensions of the work in [33] (number 9). However, some of the analysis methods are extensions of other analysis methods, which means that extensions exist between different analysis methods. For example, in analyzing the AVB flows using the CA approach by Kong *et al.* [99], they used the jitter calculations from the FA approach proposed by Benammar *et al.* [77] and also used the results obtained in the analyses of the EI approach [75] in the course of the analyses. Fig. 9 also reflects the development of each schedulability analysis method, which can help us better grasp the relevant analysis methods.

TABLE XI
SUMMARY OF THE PUBLICATIONS ACCORDING TO THE EVALUATION METHODS

| Methods | Reference to the paper | Count |
|---|---|---|
| Computational Analysis | [4, 8, 9, 11, 13, 18, 20, 21, 25, 29, 30, 34, 36, 38, 39, 43, 44, 46, 48-50, 54, 55, 59, 61, 63, 64, 66, 67, 69, 74, 78, 82, 83, 86, 88, 91, 93, 96-106] | 50 |
| Comparative Analysis | [14, 15, 37, 38, 41, 43, 47, 50-52, 55, 56, 72, 75-78, 81, 84, 85, 87, 94, 116, 119-121, 123] | 27 |
| Analysis Combined with Simulation | [2, 5-7, 10, 12, 15-17, 19, 22-24, 26-28, 31, 32, 40, 45, 52, 53, 57, 58, 62, 65, 68, 70, 71, 79, 80, 84, 89, 90, 92, 95, 109-111, 113-116, 118, 121] | 45 |

### *D. How to evaluate the result of the schedulability analysis in TSN?*

We reviewed the evaluation methods and domains in the schedulability analysis in TSN. In the following, we describe each of them.

#### *1) Evaluation Methods*

The evaluation methods used in the schedulability analysis can be mainly divided into Computational Analysis, Comparative Analysis and Analysis Combined with Simulation, as shown in Table XI.

##### *a) Computational Analysis*

Computational Analysis refers to the process of obtaining numerical results for analysis by substituting various parameters into the derived equations after the network topology and traffic have been constructed. In formal analysis methods such as Busy Period and CPA, the computational method used is often the fixed-point iteration method. For example, Thiele *et al.* [36] constructed a Quad star topology and injected CDT, general control traffic (GCT), and camera traffic (CAM) into the network, which were provided by Daimler AG. In this network, CDT is shaped by BLS. The computational conclusion shows that unshaped traffic sometimes has lower worst-case delay than CDT. Researchers also compare BLS with IEEE802.1Q and AVB. The result also shows that IEEE802.1Q had lowest worst-case delay. Unlike to formal analysis methods, many researches use Real-Time Calculus (RTC) toolbox to calculate the WCRT of NC based model, such as [50] and [41]. Shalghum *et al.* [50] first presented a flexible FWOS algorithm. To demonstrate the effectiveness of the scheduling algorithm, a simplified in-vehicle network model which is a double star is presented. Under the FWOS situation, the calculations of RTC indicate that the impact of higher-priority overlapping is larger than that of lower-priority overlapping.

##### *b) Comparative Analysis*

Some publications also provided a comparative analysis which compares the computational result with other research. For example, Zhao *et al.* [43] conducted tests on many cases, including different network topologies, combinations of different mechanisms, and selections of different parameters. Through a comparative analysis of the mechanisms, when used individually, SP and CBS demonstrate better performance. However, when multiple mechanisms are used simultaneously, the combination of TAS+ATS can combine the advantages of both and achieve even better results. The NC-based worst-case analysis of ATS was compared with the non-NC-based worst-case analysis [58], and the comparison results showed that the results calculated by NC were a little more pessimistic. Moreover, Li *et al.* [78] compared the results of their analysis with those derived from the CPA-based method proposed in [54]. The comparison revealed that, within the two considered scenarios, the ETA method provided a stricter upper bound than the CPA, with an average improvement of approximately 26% and a maximum improvement of up to 42%. Mai *et al.* [5] proposed a deep learning-based real-time Ethernet network feasibility prediction model, which has an acceleration factor ranging from 77 to 1,715 compared to schedulability analysis. However, a limitation of the deep learning-based prediction model is that its prediction accuracy only ranges from 79.3% to 90%. To address this issue, Mai *et al.* [84] proposed a prediction method that combines deep learning and schedulability analysis. When the uncertainty of the deep learning prediction is too high, the prediction is discarded and instead, the schedulability analysis is relied on. Using this method, the prediction accuracy can be improved to 99%, while the acceleration factor is 5.7 compared to traditional schedulability analysis. Comparative analysis can usually reveal the degree of pessimism of the two methods. In the work in [43], the worst-case latency calculated by the NC method and the other method were compared. For the traffic with the highest priority, the results obtained by these two methods are the same. However, as the priority decreases, the results calculated by the NC method will be more pessimistic. Cao *et al.* [75] conducted a computational comparison between the analysis method proposed in the publication and other analysis methods. The results showed that the method in the publication can provide tighter calculation results, which is attributed to the use of eligible intervals. Moreover, Kong *et al.* [99] also compared the proposed CA-based method with the FA method and evaluated the effectiveness of both approaches by presenting the CA/FA ratio, which represents the ratio of the CA results to the FA results. Notably, this publication is one of the few that demonstrates how their proposed method is more pessimistic than others.

##### *c) Analysis Combined with Simulation*

Many studies validate their hypotheses through simulations, with common simulation tools including OMNeT++ and RTaW. The schedulability analysis of the CQF with frame preemption mechanism proposed in [107] has been numerically evaluated using pyCPA and simulation validated using OMNeT++ in [159]. This approach combines the power of numerical calculations for theoretical analysis with the flexibility of simulation tools to provide a comprehensive assessment of the mechanism's performance. By utilizing pyCPA, the researchers were able to quantitatively analyze the schedulability of the proposed mechanism under various conditions and parameters. In addition, there are also some publications that use real objects for testing and verification [121]. Ren *et al.* [121] built a ring network topology and injected CDT, AVB flow and BE flow into the network. By comparing the calculated results with the measured results of real objects, it can be seen that the calculated values are basically greater than the range of measured values, and the range of difference between the maximum values of calculated and measured values is 0-300us. This is because worst-case analysis usually makes conservative assumptions, which leads to a significant overestimation of the worst-case latency and a significant underestimation of the best-case latency. Meanwhile, it is also difficult for simulations to consider the complex situations in real scenarios.

TABLE XII
SUMMARY OF THE PUBLICATIONS ACCORDING TO THE DOMAINS

| Domains | Reference to the paper | Count |
|---|---|---|
| General | [37, 39, 43, 50, 51, 53, 56, 59, 60, 63, 65-67, 69, 70, 73, 78, 81, 84-86, 89-91, 94, 96, 99, 101, 105-108, 110, 111, 117, 120, 121, 123, 128, 130, 137-140, 142-145, 148, 150, 151, 153, 154, 156-159] | 57 |
| Vehicle | [40, 41, 48, 49, 51, 54, 62-68, 71, 76, 97, 102, 113-116, 124, 125, 127-129, 131, 135, 142, 146, 147, 149] | 32 |
| Aerospace | [39, 46, 47, 57, 61, 75, 79, 80, 83, 87, 95, 102, 109, 119, 120, 122, 125, 126, 130, 132, 133, 136, 141, 153] | 24 |
| Industrial | [44, 45, 54, 57, 58, 61, 67, 74, 77, 82, 88, 92, 93, 100, 112, 113, 118, 121, 134, 141] | 20 |

*2) Domains*

TSN can provide guarantees of deterministic transmission for hard real-time systems. Therefore, they can be applied to a variety of real-time systems, such as in-vehicle networks, aerospace networks, and so on. After analyzing the schedulability, publications usually select specific scenarios to build the network to conduct related research, such as validation or extension. Therefore, we classify the evaluated scenarios according to the vehicle, aerospace, and automation domains. We must note that we categorize publications that do not specify a research scenario as general, as shown in Table XII.

*a) Vehicle*

As smart cars evolve, sensors, cameras, and Lidar increase to support advanced features such as Advanced Driver Assistance System (ADAS). These sensors introduce large amounts of data to the in-vehicle network, and some also require predictable low latency [126]. Therefore, the use of TSN in in-vehicle networks (IVNs) is a future trend, and validating the analyses against scenarios built for TSN applications on IVNs is a common approach. For example, Queck *et al.* [86] validated the results of their analysis using an example scenario of a distributed in-vehicle camera and infotainment system with a network of four cameras used for assisted driving, a Top View (TV) used to merge the video signals from the cameras, a Head Unit (HU) and a Rear Unit (RU) that displays the video signals, and a Control Unit associated with the Control Unit. Similarly, Zhao *et al.* [4] validated their analysis using a scenario where the HU and RU communicate with a front camera over two switches. Similarly, Houtan *et al.* [98] also used an in-vehicle application scenario containing control, video, audio, and diagnostic traffic.

*b) Aerospace*

Similarly, the increasing number of interconnected terminal systems in avionics equipment and the increasing amount of data that must be exchanged between them have led to an increased need for highly reliable and low-latency communications [37]. Then, TSN becomes a future solution in aerospace network communication. Zhao *et al.* [120] used the Orion Crew Exploration Vehicle (CEV) case study from [166], which has a topology consisting of 31 ESs, 15 SWs, and 39 routes connected by data flow links with a transmission rate of 1 Gbps. Control traffic and AVB traffic are in this network. This use case is also used in their other publications, such as [43], [82], and [41]. Finzi *et al.* [37] considered a Gigabit TSN-capable switch with input traffic consisting of Safety-Critical Traffic (SCT), Rate-Constrained traffic (RC), and BE in Avionics Full Duplex Switched Ethernet (AFDX).

*c) Industrial*

The Industrial Internet of Things (IIoT) envisions diverse applications, and these can place some demands on network transmission latency, especially for security-related control traffic. For example, Docquier *et al.* [17] validated the analyses using scenarios from a substation automation system (SAS) in the smart grid. Seliem *et al.* [52] built a Quality Checks After Production (QCAP) use case representing a typical industrial use case of quality checking to ensure fault-free production. The QCAP network consists of six QCAP units, ten switches and four Ethernet devices. Li *et al.* [100] validated their analyses under the case of an industrial communication system in a manufacturing plant with a hierarchical structure, where the traffic transmitted contains three types: Controller-to-Device (C2D), Device-to-Device (D2D), and Controller-to-Controller (C2C).

*3) Realistic Use Case*

In evaluating the schedulability analysis in TSN, most publications select real-world network use cases for model construction and adaptation to enhance the practicality of the research. Therefore, most publications in each of the domains incorporate realistic use cases to evaluate the effectiveness of their respective analyses. We categorize these use cases based on their scenarios, topologies, and traffic.

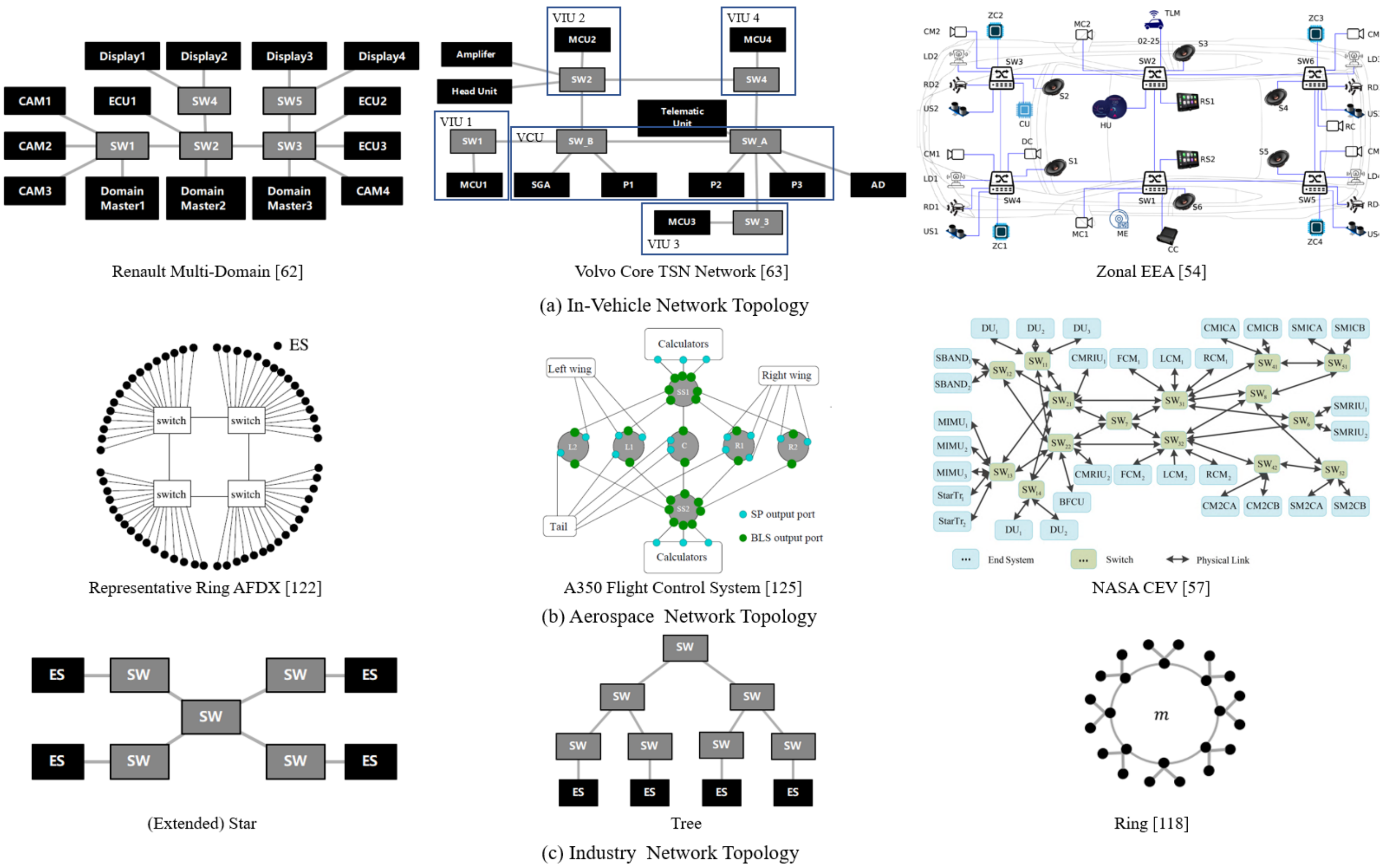

**Fig. 10.** Representative scenarios used in the publications.

TABLE XIII
SUMMARY OF THE PUBLICATIONS ACCORDING TO THE REALISTIC USE CASE SCENARIO

| Domains | Reference to the paper | Cnt |
|---|---|---|
| Vehicle | Renault Multi-Domain [63, 65, 124, 129, 144], Daimler AG Traffic [51, 66, 131], Daimler Multi-Domain [64], Volvo Backbone Network [63, 68], BMW group Multi-Domain [113], General Motors [102, 125], Unmarked Automotive Subdomain [115], Unmarked Automotive Cameras Integrated [146, 147], Unmarked ADAS Domain [71], Unmarked Entertainment Domain [142], Unmarked Multi-Domain [40, 41, 54, 67, 76, 114, 116, 127, 135] | 27 |
| Aerospace | Single-Hop AFDX Network [46, 47, 95], Representative Ring AFDX Network [47, 79, 122, 132], Unmarked AFDX Network [99, 137], A350 Flight Control [47, 136], NASA CEV [39, 57, 61, 102, 119, 120, 125, 126, 130, 133, 141, 153], Avionics Network [75], Satellite Formation Flying [83], Aviation Pilot System [87] | 24 |
| Industrial | ABB Industrial 4.0 Use Case [57, 61, 141], Industrial Production Cell [67], Common Web Communications Traffic [54], Microgrid Automation Applications [113], SAS [44], QCAP [45], Hierarchical Industrial Communication System [134, 156], 5G/6G Fronthaul Network [90], PROFINET [95], Robot Arm System [112], ALEVIN framework [100], Unmarked Factory Network [82], Unmarked Industrial Automation [74, 77, 92, 93, 118] | 21 |

*a) Scenarios*

There are several typical applications of TSN across various fields, which are frequently used in the evaluation of publications, as shown in Fig. 10.

All the publications employed in the practical scenarios for the assessment were categorized as shown in Table XIII. To enhance the clarity and reliability of the results, the sources of the scenarios referenced in the publications were labeled. In this case, "Unmarked" signifies that the publication does not specify the source manufacturer associated with the scenario or the commonly recognized name under which the scenario is known. For the scenarios where the publication does not explicitly specify a use case, they are not included in the count. Additionally, some publications employed multiple evaluation scenarios, all of which have been included in the summary of our classification.

In the automotive domain, TSN are commonly employed in automotive backbones to transport traffic across various functional domains, such as ADAS and Entertainment System.

For example, Mai *et al.* [62, 63, 65] employed a multi-domain use case involving the co-transmission of control system, ADAS, and entertainment system traffic within the in-vehicle network provided by Renault. Thiele *et al.* [51, 66, 131] constructed realistic evaluation use cases by deploying Daimler AG-provided traffic within the network topologies they designed. Thomas *et al.* [68] employed the Volvo core TSN network to construct an evaluation network within a software simulation.

Similarly, the scenarios employed to evaluate schedulability analyses in the aerospace domain encompass a variety of use cases. With the adoption of AFDX in avionics, both single-hop AFDX use case [46, 95] and representative ring AFDX network use case [79, 122, 132] were employed in the evaluation process. The evaluation of the analyses was also conducted by Finzi *et al.* [47, 136], utilizing Flight A305 control traffic within the corresponding network architecture. It is noteworthy that the most widely utilized use case in the aerospace domain is, in fact, the CEV use case from NASA, employed by Zhao et al. [39, 120, 125, 126, 130, 133, 153].

TABLE XIV
SUMMARY OF THE PUBLICATIONS ACCORDING TO THE EVALUATION TOPOLOGIES

| Domains | Reference to the paper | Count |
|---|---|---|
| (extended) star | [44, 46-48, 83, 95, 112, 121] | 8 |
| line | [40, 41, 49, 51, 54, 58, 66, 75, 76, 80, 82, 87, 88, 92, 93, 109, 113, 114, 127, 131, 134, 135, 142, 146, 147] | 25 |
| Tree | [51, 67, 74] | 3 |
| Ring | [45, 47, 71, 77, 79, 102, 118, 122, 132, 136] | 10 |
| Mesh | [39, 57, 61, 100, 102, 119, 120, 125, 126, 130, 133, 141, 153] | 13 |
| Domain EEA | [62-65, 67, 113, 116, 118, 124, 129] | 10 |
| Zonal EEA | [54, 63, 68, 149] | 4 |

TABLE XV
SUMMARY OF THE PUBLICATIONS ACCORDING TO THE EVALUATION TRAFFIC CLASSES

| Domains | Reference to the paper | Count |
|---|---|---|
| CDT | [39-41, 44-49, 51, 54, 57, 58, 61-68, 71, 74-77, 79, 82, 83, 87, 88, 92, 93, 95, 97, 100, 102, 109, 112-114, 116, 118-122, 124-127, 129-136, 147, 149, 153] | 62 |
| AVB | [39-41, 44-49, 51, 54, 57, 58, 61-68, 71, 75, 76, 79, 87, 93, 95, 97, 100, 102, 109, 113, 114, 116, 118-120, 122, 124-127, 129-136, 141, 142, 146, 147, 149, 153] | 57 |
| BE | [40, 44-49, 54, 58, 62, 63, 65, 67, 68, 74, 75, 77, 79, 82, 87, 93, 95, 100, 102, 112, 116, 118, 121, 122, 124, 127, 129, 132, 134, 136, 149] | 36 |

In the industrial domain, the adoption of TSN is also increasing. With the advent of Industry 4.0, relevant use cases proposed by the ABB group [141, 145, 147] have been developed. Zhang *et al.* [134] and Li *et al.* [58] have also explored the impact of schedulability analysis within industrial communication networks with a hierarchical structure. Shen *et al.* [112] employed the Robot Arm System as a scenario to construct an evaluation environment for their schedulability analysis.

*b) Topology*

The various scenarios mentioned above differ in characteristics such as the scale of their application, traffic attributes, and implemented functionalities, resulting in a broader range of topologies. In addition, Fig. 10 illustrates several representative topologies.

In the vehicle domain, the topologies of in-vehicle networks used in current publications can be divided into two categories from the point of view of automotive Electronic and Electrical Architectures (EEA): distributed, functional domain and zonal architecture. Vehicle functional domains are groupings of systems within a vehicle that are responsible for specific functions or tasks. Some Original Equipment Manufacturer (OEM) uses the in-vehicle backbone network to transport multiple functional domain traffic. For example, Renault offered an in-vehicle network architecture for traffic transport in the control, ADAS, and entertainment domains [63, 65, 124, 129, 144], as shown in Fig. 10(a). Daimler [64] and BMW group [113] have also built similar backbone network architectures for multi-domain traffic transmission. In addition, as in-vehicle network architectures evolve towards centralization, zonal architectures are gradually being taken more seriously by more OEMs like Volvo [63, 68]. A zonal architecture typically consists of multiple zonal controllers and a central computing unit. Each zonal controller is responsible for managing key functions in the respective zone. Moreover, some publications use distributed network topologies of their design, such as line [66, 131] and tree [51], paired with real vehicle traffic for evaluation.

In the aerospace domain, network architectures are often complex. The NASA CEVs uses dozens of switching devices to form a complex mesh-shaped topology, as shown in Fig. 10(b). Meanwhile, to ensure communication reliability, ring topology is also widely employed in aerospace. A350 Flight Control System achieves communication through network topologies featuring multiple rings [47, 136]. Some publications [47, 79, 122, 132] have been evaluated using the Representative Ring AFDX Network consisting of four switches and several end stations, as shown in Fig. 10(b).

In industry domain, star, extended star, tree, and linear topologies are often chosen for transmission for ease of expansion and maintenance, as shown in Fig. 10(c). For example, Bello *et al.* [67] and Benammar et al. [113] used the linear topology in the Industrial Production Cell and Microgrid Automation Applications as an architecture for evaluating the network. Docquier et al. [44] used an extended star topology in SAS to evaluate the analysis.

Our categorization of publications according to the topologies of the realistic use cases is shown in Table XIV.

*c) Traffic*

Similar to the above categorization of the traffic analyzed, we classified the traffic injected in these cases into three types, CDT, AVB, and BE, and categorized the publications as shown in Table XV. Some of the publications were injected into the network using OEM-provided traffic. For example, Thiele *et al.* [51, 66, 131] injected traffic provided by Daimler AG into the network, which included attribute definitions for CDT and AVB traffic. Since the Volvo TSN network use case contains control, ADAS, and entertainment domains, the traffic under this use case contains CDT, AVB, and BE traffic [63, 68].

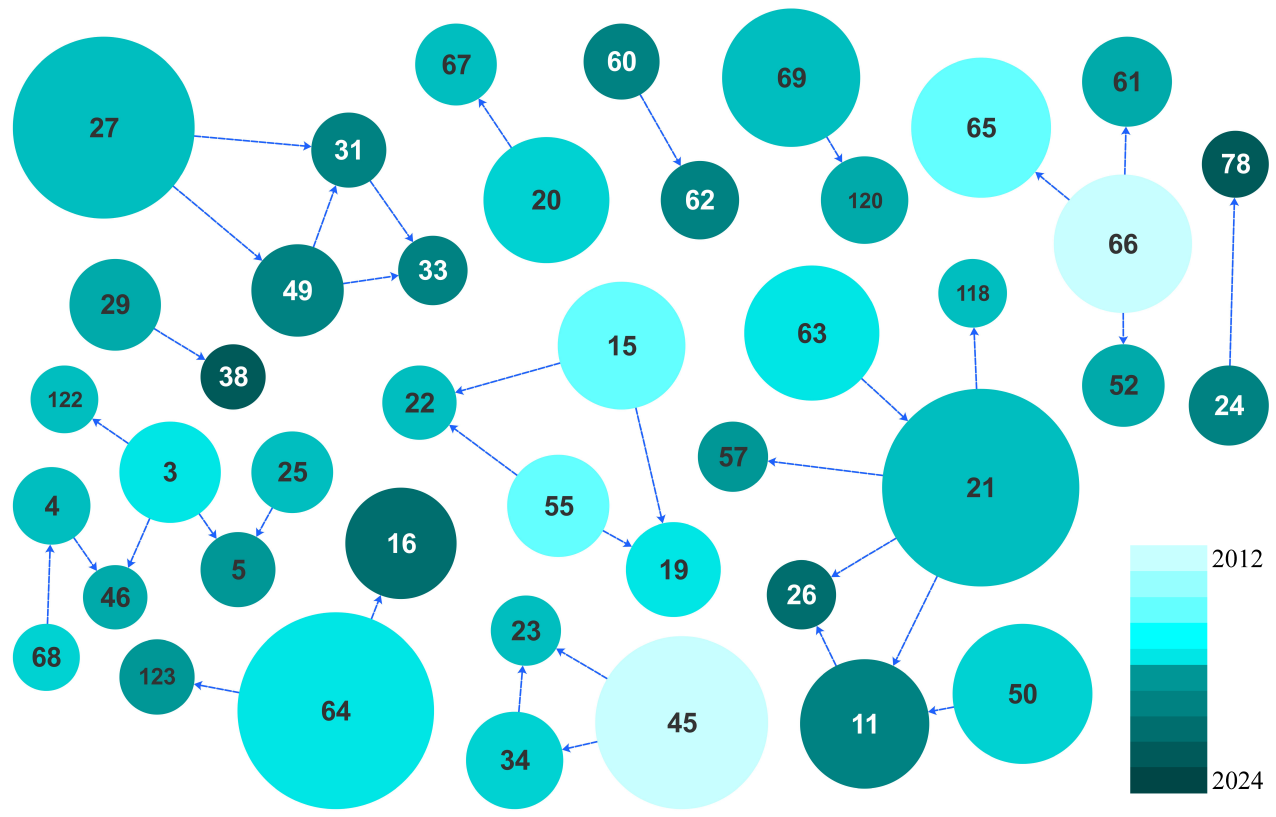

**Fig. 11.** Comparative relationship between the publications.

TABLE XVI
SUMMARY OF THE SIMULATION PLATFORM OF THE PUBLICATIONS

| Simulation Platform | Reference to the paper | Count |
|---|---|---|
| OMNeT++ /OMNEST | [10, 15, 17, 19, 27, 28, 40, 45, 52, 53, 57, 58, 65, 68, 70, 71, 79, 89, 92, 113-116] | 23 |
| RTaW-Pegase | [5, 22, 84, 95] | 4 |
| NS-3 | [62, 110] | 2 |
| Python | [12, 90] | 2 |
| Hardware | [7, 24, 31] | 3 |
| OPNET | [121] | 1 |
| Vector CANoe | [2] | 1 |
| Other | [23, 26, 32] | 3 |
| Unknown | [6, 16, 80, 109, 111, 118] | 6 |

*4) Insights from RQ4*

Summarizing the above classification and combining it with Table XI, most current research still prioritizes validating the results of schedulability analysis using Computational Analysis, followed by Comparative Analysis and Analysis Combined with Simulation. Moreover, from Table XII, the current scenarios selected for evaluation in most publications are generic, followed by in-vehicle network scenarios.

For publications using comparative analysis, we plotted their relationship to the publications they were compared to, as shown in Fig. 11. The circles in this figure represent the same meaning as in Fig. 2 and 9. In the figure, the publications at the head of the arrow cite the publications at the arrow's tail for comparison. As the figure shows, some publications, such as [36], [120], and [86], are often used for comparison. Combined with Table VIII, we can see that comparisons can occur between different analysis methods. For example, Cao et al. used EI approach to analyze the schedulability of AVB traffic under the CBS mechanism [75, 76]. They compared their analysis with those of [74] using the GBPA approach and [83] using the CPA approach and concluded that their proposed method provides tighter latency upper bounds. The results of these comparisons can be used as a reference for analyzing the effects between approaches. In addition, publications applying the same analysis method may also be compared. As an example, Zhao et al. [56] derived a delay upper bound for the CDT traffic under TAS using the NC approach and compared their results with [48], which also used NC. This also shows that the level of detail of the analysis can affect the effectiveness of the same schedulability analysis method. It also means we cannot simply conclude that one method is better than the other by comparing these results. It is important to note that the publication that uses other approaches to reproduce results for comparative analyses [15] is also classified under this category, but is not represented in the figure.

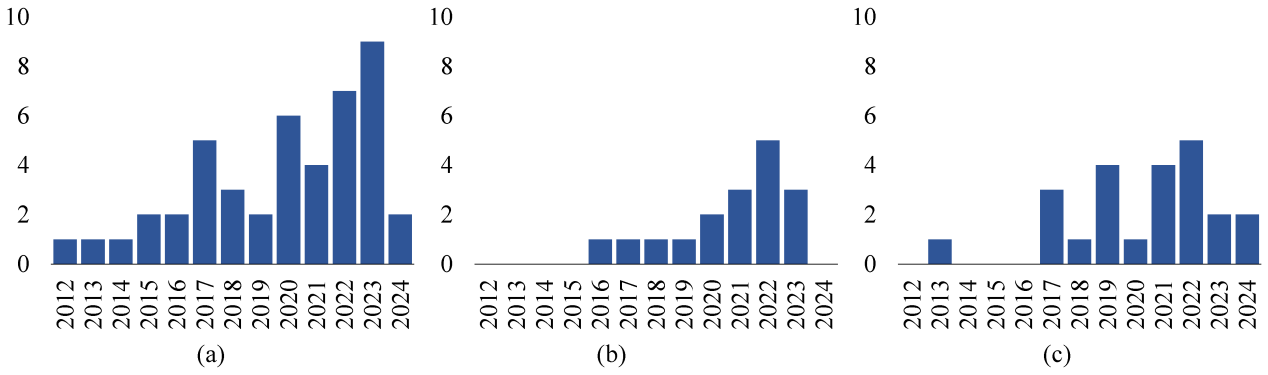

**Fig. 12.** Trend of analysis objectives: (a) Mechanism Evaluation, (b) Constraint Construction, and (c) Configuration Viability.

We summarized the realistic use cases used by the publication to validate the simulation and the scenarios in which they are applied. As can be seen in Table XIII and Table XIV, the publications mainly use linear topologies, and the most used scenario is multi-domain traffic transport in the automotive domain, where some of the publications also use realistic solutions given by OEMs.

We categorize publications that analyze combined simulation according to their simulation platform, as shown in Table XVI. From the Table, we can see that the simulation validation in most research is based on OMNeT++/OMNEST platform, which is event-based. Note that OMNeT++ is an open-source version, and OMNEST is the commercial version. Thus, OMNeT++ is often used in academic research. Additionally, OMNeT++ simulations provided a practical evaluation of the mechanism's behavior in a simulated environment, allowing for the identification of potential issues and optimization opportunities. RTaW-Pegase is a time-precise simulation that only offers a commercial version, which is less widely used than OMNeT++. The table also shows that OPNET [121] and Vector CANoe [2] are used for simulation, but they are used very infrequently.

## V. DISCUSSION

Over the past decade or so, there has been a proliferation of research on TSN schedulability analysis. This section will focus on the trends we have observed when conducting SLR and the challenges that the community should still strive to address.

### *A. Trend Analysis*

Fig. 12 shows how the purpose of schedulability analysis in publications has changed over time. Since one of the aims of all the publications related to schedulability analysis is to obtain tight boundaries, it follows the same trend as the overall publication trend, as shown in Fig. 5. Overall, most schedulability analyses in TSN aim at mechanism evaluation, and every year, there are studies with it as their purpose. Since 2016, publications on the purpose of configuration viability have gradually appeared. However, although fewer publications aimed at constraint construction, they have accounted for a larger share in recent years. Therefore, with the development of schedulability analysis in TSN, there may be more publications that use the results for constraint construction in scheduling or routing algorithms, and provide recommendations for parameter configuration. Moreover, evaluating single or multiple scheduling mechanisms and providing tighter latency upper bounds through schedulability analyses will also remain a focus.

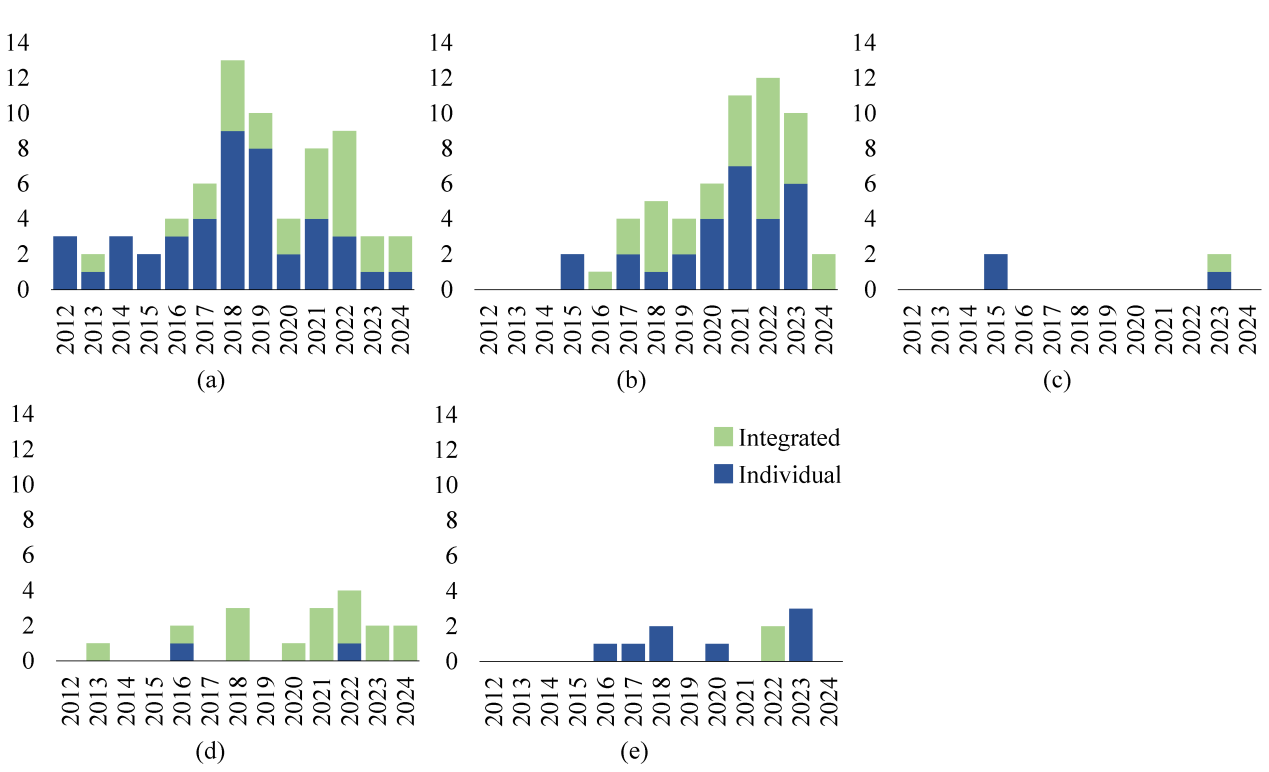

**Fig. 13.** Trend of scheduling mechanisms analyzed: (a) CBS, (b) TAS, (c) CQF, (d) Preemption, and (e) ATS.

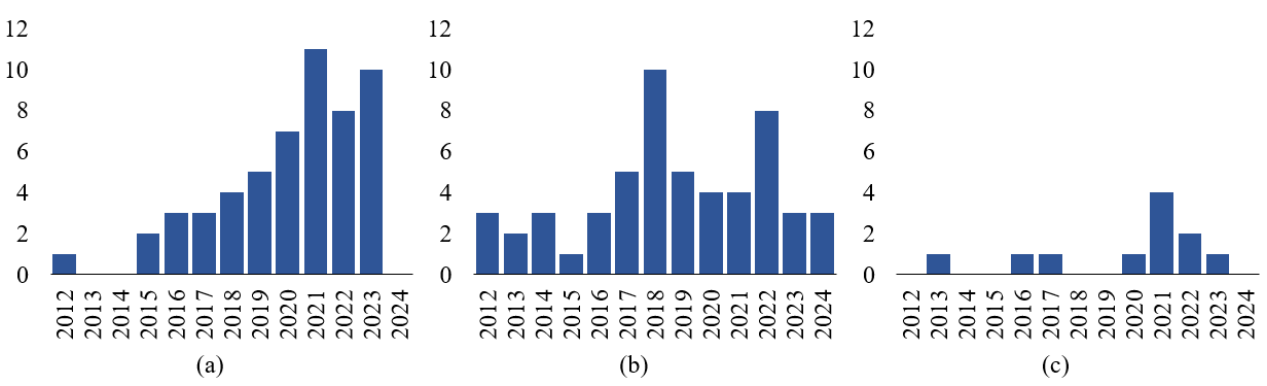

**Fig. 14.** Trend of traffic classes analyzed: (a) CDT, (b) AVB, and (c) BE.

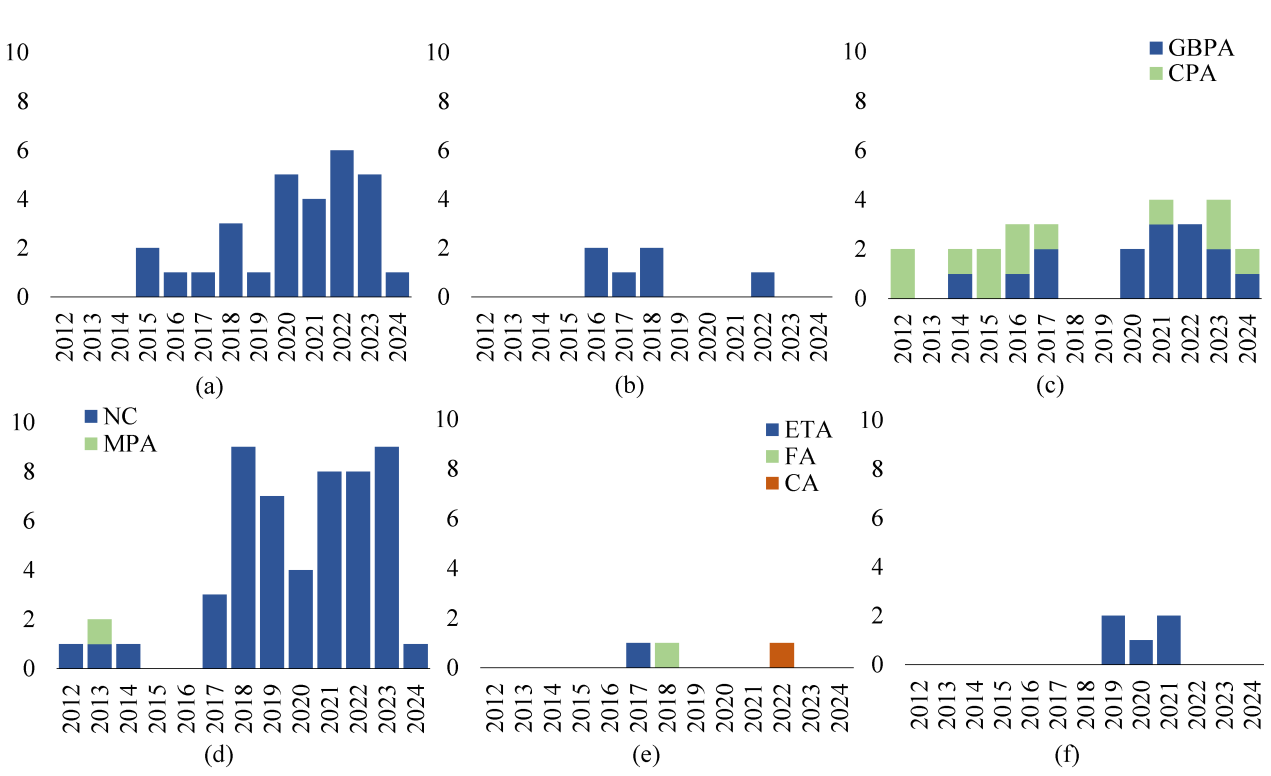

**Fig. 15.** Trend of analysis methods: (a) FTA, (b) EI, (c) BPA (GBPA and CPA), (d) NC (NC and MPA), (e) TA (ETA, FA, and CA), and (f) IA.

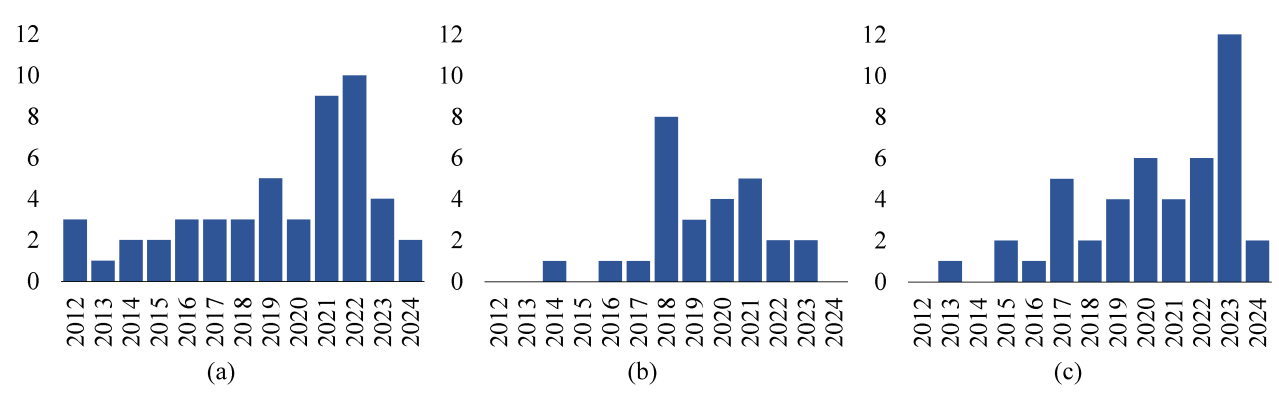

**Fig. 16.** Trend of evaluate methods: (a) Computational Analysis, (b) Comparative Analysis, (c) Analysis Combined with Simulation.

The trend in the analyzed scheduling mechanisms shown in Fig. 13 demonstrates the large number of publications that target CBS for analysis each year. The percentage of publications that address the use of CBS in combination with other scheduling mechanisms is also increasing. Although publications for TAS did not appear until 2015, the trend of publications related to it is similar to that of CBS. Although the number of publications on preemption, CQF, and ATS is low, publications have analyzed them in recent years. The above analysis shows that the number of publications on preemption, ATS, and CQF mechanisms may gradually increase. In addition, schedulability analyses using multiple scheduling mechanisms will also be a trend in the future.

Fig. 14 illustrates the variation of the analyzed traffic classes. The analyses for AVB traffic have been dominant. This trend is similar to that of the study of the CBS mechanism because the main target of the CBS mechanism is AVB traffic. This is because more publications have studied BLS, which targets CDT, during these years. However, as mentioned above, because of the similarity of its mechanism to CBS, we classified it under the category of CBS. Most publications focus on WCRT for CDT and AVB traffic, but the research on BE traffic has also gradually increased in recent years. In future schedulability analyses, in scenarios with multiple scheduling mechanisms, the publication may further optimize the analysis of delay upper bounds for BE traffic while maintaining the focus on CDT and AVB traffic.

The trend of analysis methods shown in Fig. 15 demonstrates that the NC approach is the most used method for schedulability analysis in TSN. Although its use was limited in previous years, it increased significantly in 2018. As the figure shows, the CPA approach was used more frequently in the first six years and has been used in several publications in recent years. Although the GBPA approach is not used as often as the NC approach, they have also appeared in publications in recent years. Recently, there have been a large number of publications using the FTA approach almost every year. Over time, the EI, TA, and IA approaches have been refined regarding effectiveness and generality, but only some relevant publications exist because they appeared later.

Fig. 16 illustrates the evolution of the methodology for evaluating the results of schedulability analyses in TSN over time. Computational analyses are the most used assessment method in publications every year. Overall, the number of uses of the method grows roughly over time and peaks in 2022. With the development of schedulability analysis in TSN, more publications have demonstrated their optimization by comparing them with previous results, as shown in Fig. 16(b). In addition, with the development of simulation technology, more and more work tends to use simulation work to verify and analyze network performance. Therefore, some publications have incorporated simulation to validate the results of recent developments in software simulation techniques. As a result, several publications have used simulation techniques to validate recent developments in schedulability analysis. The number of publications in this category has increased significantly in recent years and is likely to increase.

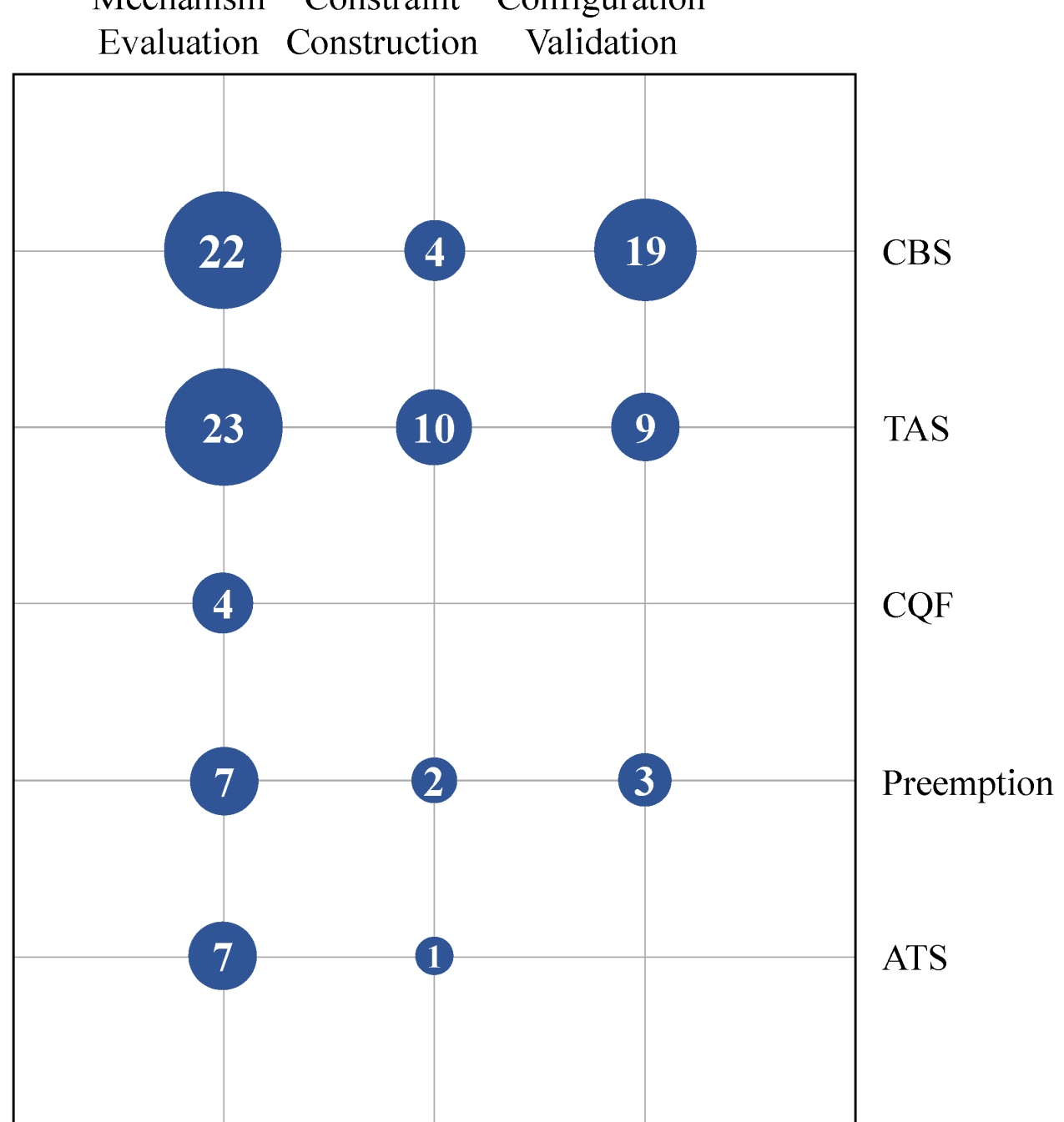

**Fig. 17.** Relationship between Analysis Objectives and Scheduling Mechanisms in the publications.

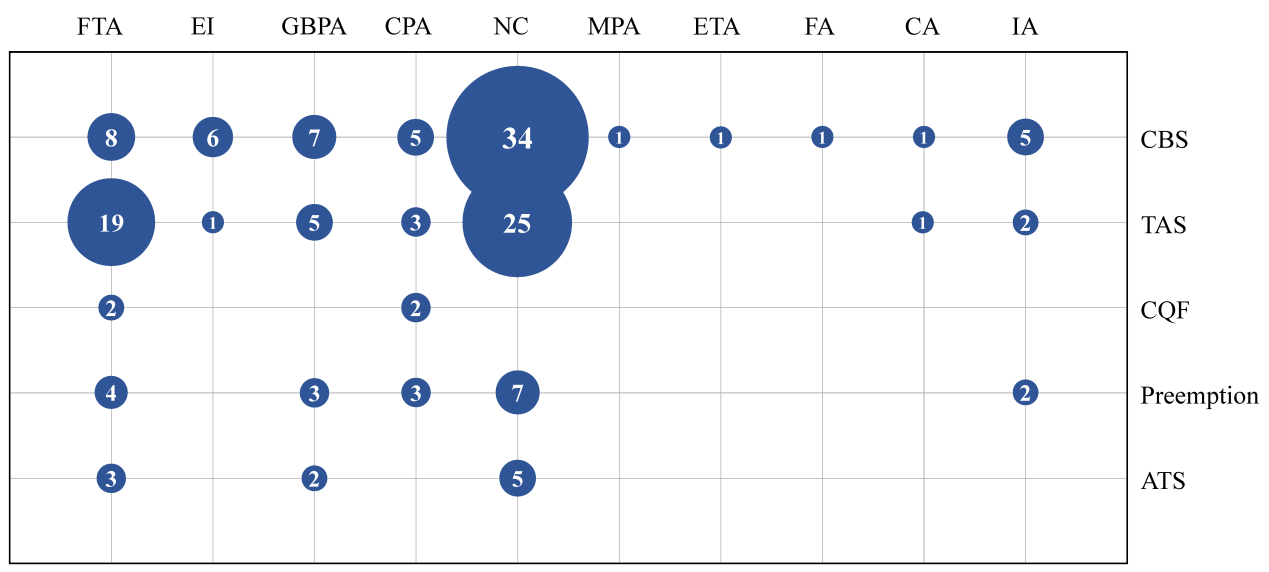

**Fig. 18.** Relationship between Analysis Methods and Scheduling Mechanisms in the publications.

*B. Analysis of the Scheduling Mechanisms*

We used the bubble diagrams to analyze the relationship between two different categories, where the size of the bubbles corresponded to the number of publications involving pairs of categories that intersected with each other.

Fig. 17 shows the relationship between analysis objectives and scheduling mechanisms. The horizontal axis of Fig. 17 represents the publication's analysis objectives, and the vertical axis represents the scheduling mechanisms for the publications. As can be seen from the figure, the current publications of the schedulability analysis of CBS and TAS focus on tight boundaries, mechanism evaluation, and configuration viability, with relatively few publications on constraint construction and validation & implementation. The research objectives of current publications related to CQF mainly focused on mechanism evaluation, and there need to be more optimization studies for tighter E2E latency and backlog boundaries and studies aimed at establishing constraints in the solution process. Since preemption is usually used with other scheduling mechanisms, the number of publications related to preemption is high compared to CQF and ATS and covers all analysis purposes. However, the number of publications related to it still needs to be more extensive compared to CBS and TAS. In addition, configuration feasibility studies for the CQF and ATS mechanisms currently also need to be improved. Although the CQF mechanism can provide an upper bound on latency (related to the number of hops and the cycle time), a tight latency bound provides the network designer with a broader scope of involvement and reduces the waste of network resources. Moreover, the study of bounds for the backlog is also necessary for the CQF mechanism.

Similarly, we explored the relationship between schedulability analysis methods and scheduling mechanisms, as shown in Fig. 18. As can be seen from the figure, publications analyzed for CBS used all the methods analyzed, with the NC approach having the highest number at 26. Even though there were also several publications analyzed for TAS, some more recently proposed methods were not used, such as ETA and FA. Only the FTA and CPA methods have been used in current studies for CQF, and they are both small in number, with ATS being similar. In addition, from the point of view of the analysis method, Fig. 18 reflects that the only method that analyses all the scheduling mechanisms is FTA, which is because FTA's approach does not have mandatory rules. An analysis result can be obtained only through a formal expression, which also leads to a need for more guarantees on the quality of the results (which will be analyzed in the following section). The EI, CA, and IA approaches are currently only applied to the schedulability analysis of CBS and TAS (the IA approach includes preemption), and their application to other mechanisms remains to be explored. For the more commonly used methods, the GBPA and NC approaches have not been applied to CQF, and the CPA approach has not been applied to ATS. In addition, analysis approaches such as ETA, FA, and CA have the potential to be applied in other mechanisms in TSN, as they performed well in the schedulability analysis of CBS. It is important to note that using different methods for scheduling mechanisms may increase the likelihood of lowering the latency bound.

*C. Analysis of the Schedulability Analysis*

*1) Analysis Extensiveness*

As shown in Fig. 19, we use a bubble chart to show the relationship between the schedulability analysis methods (vertical axis) and the blocking considerations (horizontal axis). In this figure, we only show the analysis approaches used in the publications that lacked blocking considerations (FTA, EI, GBPA, and NC), indicating that publications of other approaches fully consider all types of blocking.

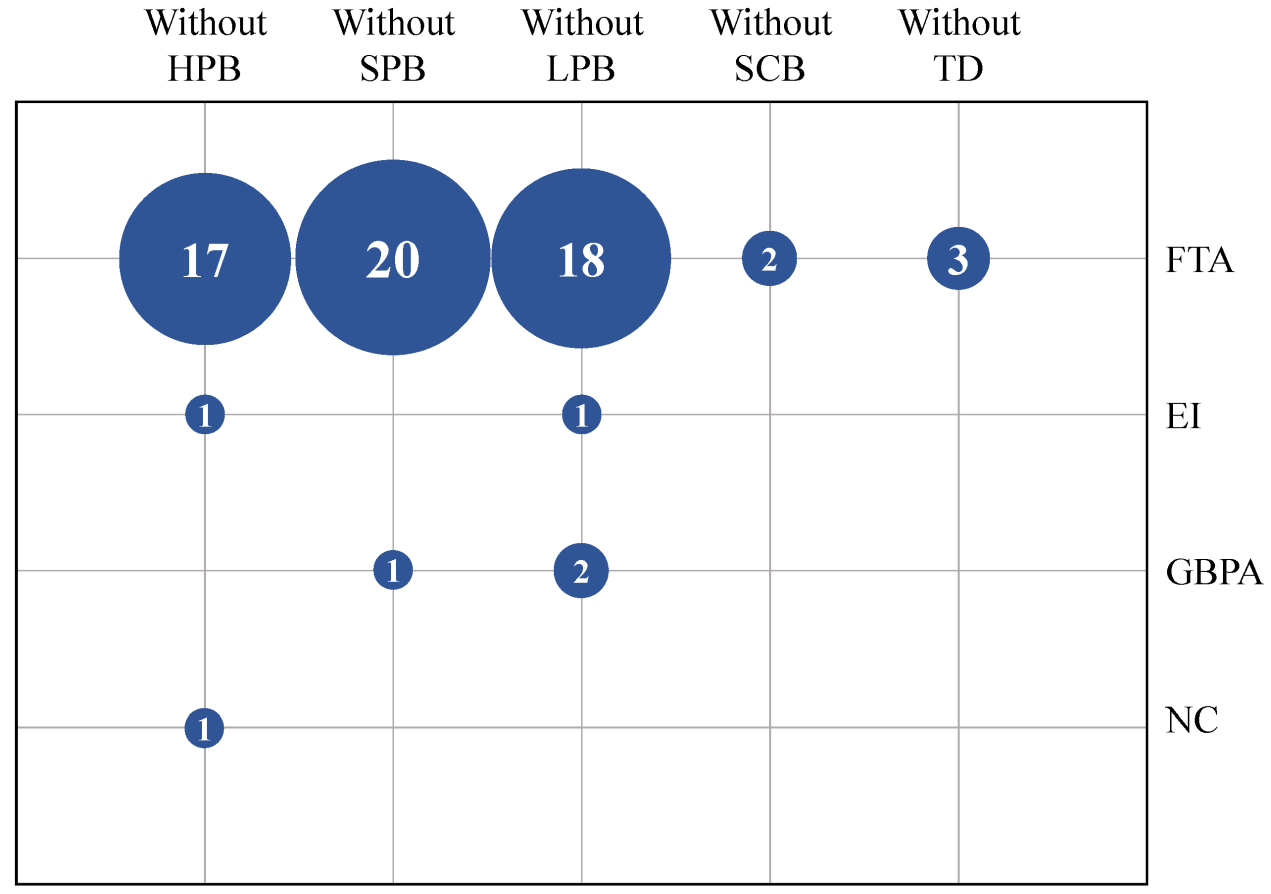

**Fig. 19.** Relationship between the Blocking Consideration and Analysis Methods in the publications.

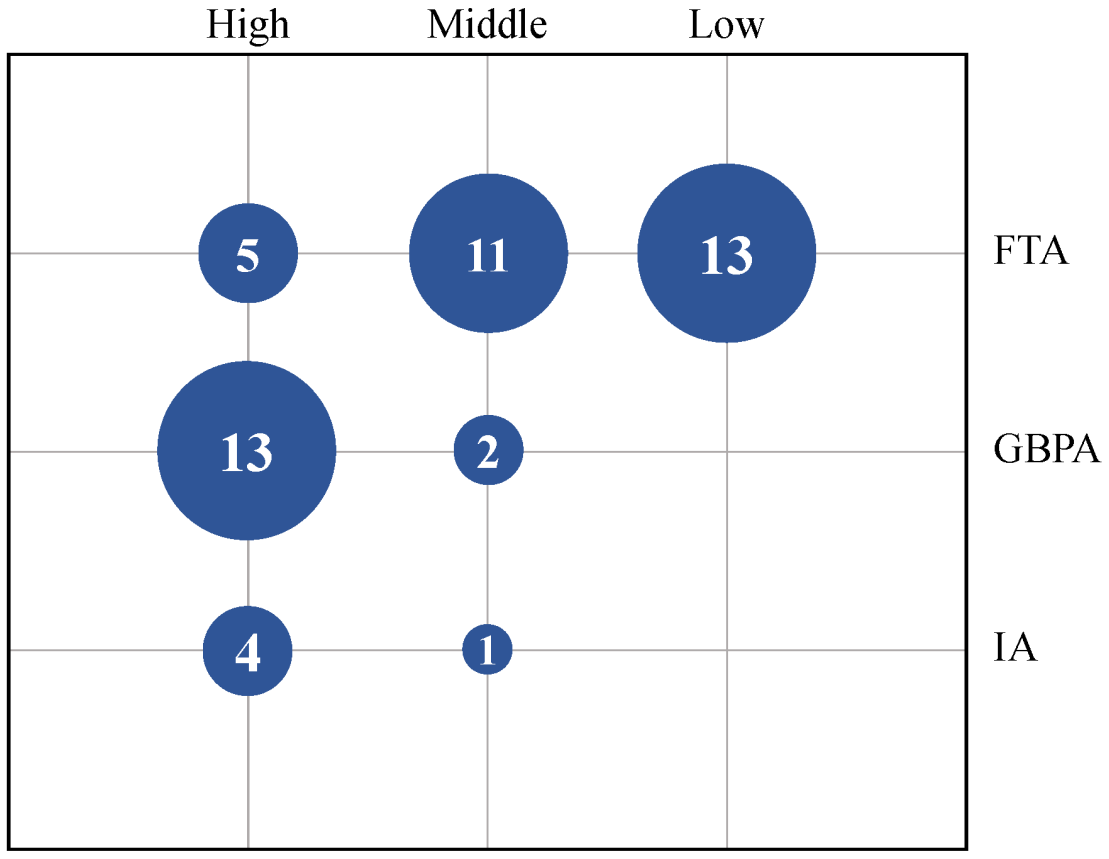

**Fig. 20.** Relationship between the Level of detail in Modeling and Analysis Methods in the publications.

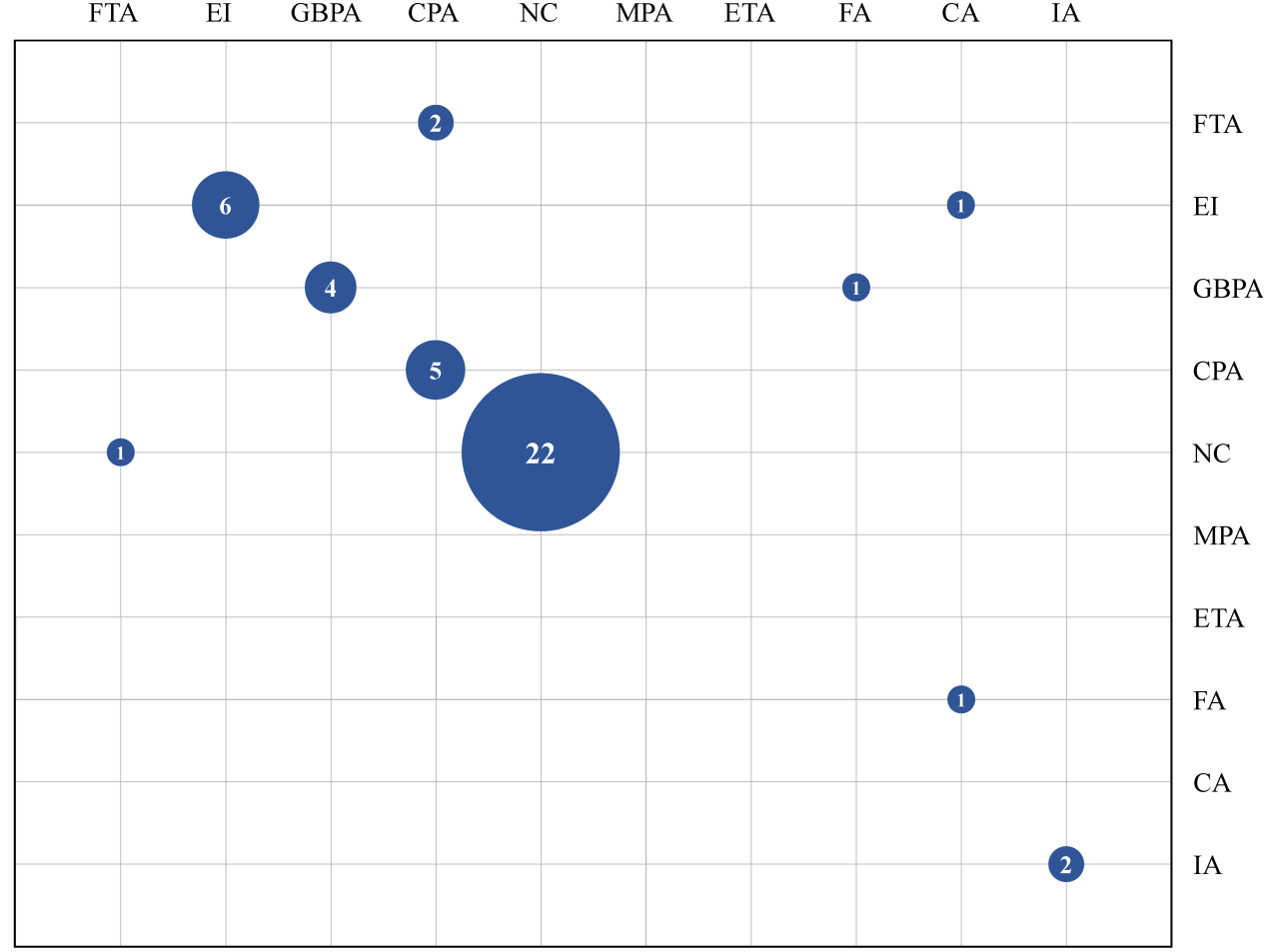

**Fig. 21.** Extended relationship between the analysis methods.

As can be seen from the figure, the publications using the FTA approach needed more blocking considerations by the largest percentage, with up to seven of them not refining the blocking of frames of the same priority during transmission. Some publications do not consider the transmission delay of the frames because the FTA approach does not consider the rules of blocking during the analysis, which leads to the poor quality of its results. In Fig. 19, one of the publications of the EI approach does not consider lower-priority blocking because it assigns the lowest priority to the traffic being analyzed during the analysis. Although this reduces generality, it is related to the scenario building and is not flawed in the method. The situation is similar for GBPA and NC in Fig. 19. In addition, one publication used the EI approach that did not consider the co-existence of higher and lower-priority blocking (which we classify in this SLR as not considering HPB). However, this issue was resolved in a subsequent publication.

We have also represented the relationship between the Level of Detail in Modeling and Analysis Methods, as shown in Fig. 20. Similarly, we only show the analysis methods (FTA, GBPA, and IA) associated with publications rated Middle and Low, which implies that the level of detail ratings of publications associated with the other methods are High.

As can be seen from the figure, the level of detail in modeling of the publications using the FTA approach is mostly low. Most of publications using the FTA approach for modeling streams and scheduling mechanisms cannot cover detailed stream properties and the parameters of scheduling mechanisms. Although some publications include detailed parameters in the FTA approach modeling process, they are fewer. Conversely, the GBPA and IA approaches were rated High in most numbers. However, some publications using the GBPA approach should have considered the influence of the stream parameters when analyzing the WCRT (no stream modeling). Therefore, they are rated as Middle.

As analyzed above, we can conclude that although we cannot judge the analysis methods by their results, as this is related to the blocking consideration and level of detail in the construction of the model, some of the analysis methods can provide a basic analytical framework to enable more detailed analysis. However, as mentioned before, the FTA approach does not have constraint rules during the analysis, leading to mixed results in quality.

*2) Extension of Analysis Methods*

We also used a bubble diagram to represent the expanded relationships between the analysis methods, as shown in Fig. 21. We use the horizontal axis to represent the methods used for the extended publications and the vertical axis to represent the methods used for the extended base publications. For example, the number 2 at the intersection of CPA on the horizontal axis and FTA on the vertical axis indicates that two publications are optimized and expanded studies based on the publications of the FTA approach, and the method used for the extended publications is CPA. Fig. 21 describes Fig. 9 at the level of the analysis method, with the numbers in Fig. 21 corresponding to the number of arrows in Fig. 9.

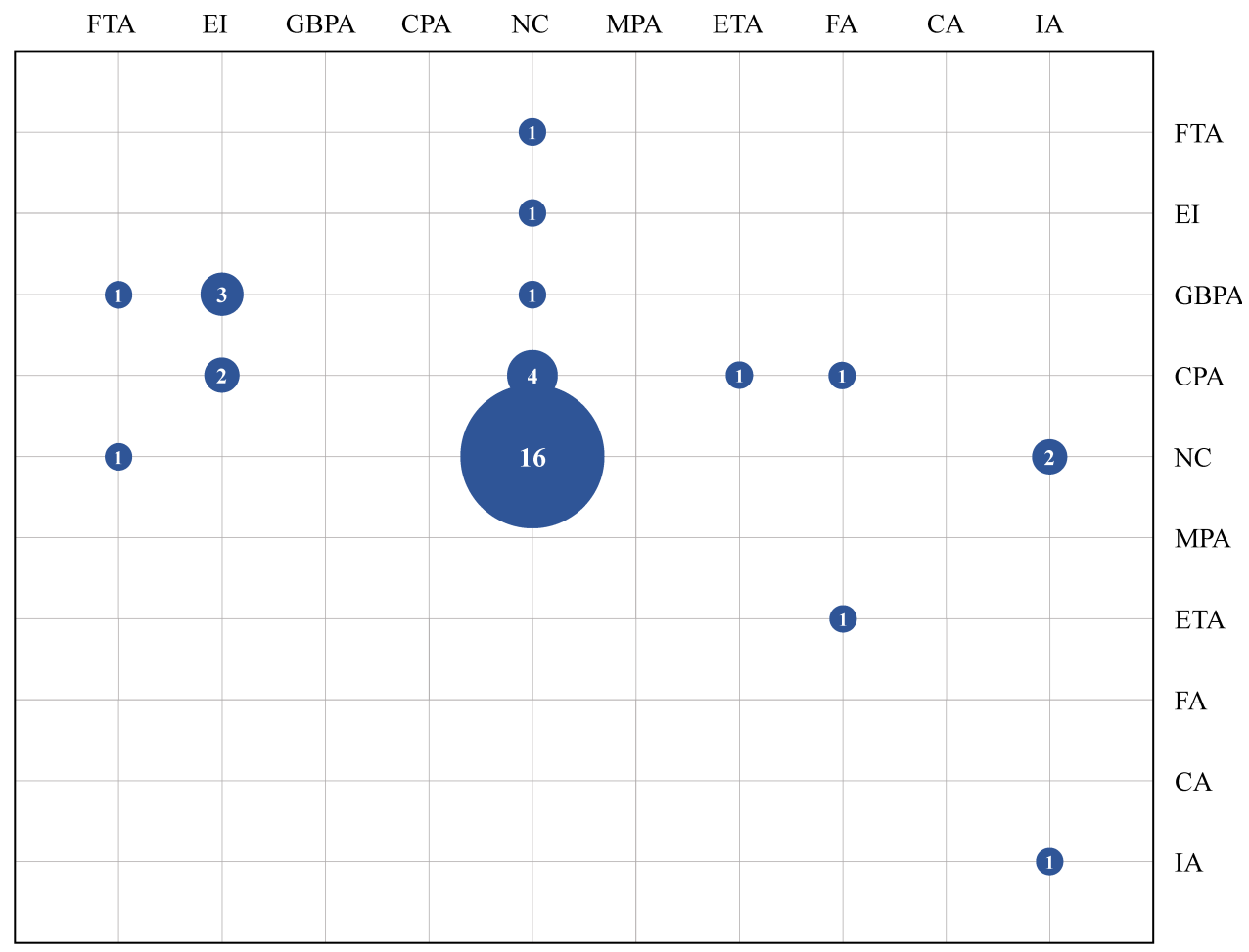

**Fig. 22.** Comparative relationship between the analysis methods.

As seen from the figure, most extended methods are the same. For example, the publications that were extended using the EI, GBPA, NC, and IA approaches are all based on those that used these methods. In addition to the two publications using the CPA approach mentioned above, the figure shows that the publications using the FA approach were obtained by expanding based on the one using the GBPA approach. Further, the publication using the CA approach was obtained by expanding on the publications using the EI and FA approaches, which is why it is called combinable.

Fig. 21 allows us to analyze the extension relationship at the level of schedulability analysis methods. It can be seen from the figure that most of the current extensions are based on the NC approach, which also shows that the NC approach is more widely used and has high scalability, which means that it is more flexible to adapt to different scenarios and needs of the analysis. Similarly, the EI, GBPA, and CPA approaches are also scalable.

*3) Comparison of Analysis Methods*

Similarly, we show the comparative relationship between different schedulability analysis methods, as shown in Fig. 22. In this figure, the horizontal axis represents the analysis methods used by the publication that used the comparison method, and the vertical axis represents the analysis methods used by the publications that were used for the comparison. For example, the number 1 in the intersection of NC on the horizontal axis and FTA on the vertical axis indicates that one publication compared its results using the NC approach with other publications using the FTA approach. It is important to note that the summary shows that all publications using the comparison method have better results than those being compared. This also means that the publication using the NC approach in the example above produced tighter latency bounds than the FTA approach. Similarly, Fig. 22 can be seen as depicting Fig. 11 at the level of the analytical approach, with the numbers in Fig. 22 corresponding to the number of arrows in Fig. 11.

The largest number of comparisons is between publications using the NC approach, as shown in Fig. 22, which corroborates the content in Fig. 21. This also shows that optimization for schedulability analysis for the NC approach is an important research direction. As seen in the figure, from the vertical axis, the NC approach is used for comparison with several other methods, such as the FTA, EI, and CPA approaches. In addition to the GBPA, CPA, MPA, and CA approaches, publications using other analysis methods have been evaluated comparatively. In addition to this, on the horizontal axis, the most used analysis method for comparison is the NC approach, followed by the CPA approach. All other analysis methods are used for comparative assessment except for the MPA, FA, and CA approaches. It is important to note that although the results derived from the CPA approach are often used for comparison while not actively compared with other methods, this does not mean that the approach produces inferior results to the other approaches, but rather indicates that there is a need for more optimization studies using the CPA approach. The GBPA and CA approaches are similar to this.

*D. Open Issues and Future Work*

Based on the publications, we have summarized the future work and challenges proposed by the authors and combined this with the above analysis to come up with the following open issues and future work.

*1) More/Combined Mechanisms Support*

The above analysis shows that the mechanisms analyzed in the current schedulability analysis in TSN are mainly focused on the CBS and TAS mechanisms. At the same time, less research has been carried out on scheduling mechanisms such as CQF, preemption, and ATS, which has led to several publications pointing out that analyses of other scheduling mechanisms are necessary for future work. In addition, scenarios for the combined use of scheduling mechanisms are unavoidable, as future networks will be more complex, and various classes of traffic with different latency requirements may be transmitted in the network. From the above analyses, the current scenarios of the combined use of scheduling mechanisms are mainly related to CBS, TAS, and preemption mechanisms, and more combinations of other mechanisms have yet to be explored.

*2) More Traffic Classes Support*

The current traffic classes for schedulability analysis are mainly CDT and AVB traffic. However, the current analyses of different traffic classes under different scheduling mechanisms or analysis methods need to be revised. For example, Zhao *et al.* [15] proposed an FTA approach to obtain the WCRT for AVB traffic, but they also pointed out that subsequent work needs to apply the method to CDT traffic. In addition, although some current publications do analyses for BE traffic, the overall number is small, and the studies for BE traffic need to be more comprehensive at the level of scheduling mechanisms and research methods. Although there may not be strict latency requirements for BE traffic, it is usually expected to achieve an acceptable Quality of Service (QoS) level, especially when supporting legacy traffic [73].

Meanwhile, as mentioned above, when the network is complex, the scheduling mechanism combined with the situation may put different demands on different traffic types. Hence, analyzing latency upper bounds for more traffic classes is also necessary.

*3) Optimization of Analysis Methods*

The above analyses show that various analytical methods are also being optimized. For example, although the NC approach is the most widely used analysis method, its optimizations are also the highest. In addition, a growing number of publications also focus on providing tighter upper bounds through other methods, such as trajectory method-derived methods (the ETA and FA approaches) and intelligent algorithms. This also shows the potential for better results through methods other than traditional schedulability analysis (the NC and BPA approaches). The tighter upper bounds of latency and backlog can provide us with a more reliable design basis and, at the same time, reduce the waste of network resources. Therefore, optimizing the analysis methods to provide tighter delay upper bounds is an essential direction for future work.

*4) Validation of Analysis*

The validation of the results of schedulability analyses currently takes the form of analyses of computational results. With the expansion of schedulability analyses in TSN and the development of simulation platforms, future validation of analysis results may favor comparisons with other work and simulation results. Meanwhile, a reliable simulation platform and tools for calculating schedulability analysis results are also essential. Although this SLR does not provide a detailed summary of these tools and platforms, developing reliable simulation platforms and calculation tools is also a trend for future research.

## VI. Threats to Validity

To demonstrate the quality of our research, we list the threats that this SLR may face and how we have mitigated them.

### *A. Omission of Publications*

There may be instances where publications are missed during the literature collection process. To mitigate this, we selected the four most commonly used and complete databases in computer science-related areas: the ACM Digital Library, the IEEE Xplore Digital Library, Scopus, and Web of Science. In addition to this, we used a grey literature search method on the ArXiv E-print Repository and Google search engines. Following the automated search, we implemented a forward and backward snowballing strategy to ensure that all possible publications were covered. In addition, we disregarded publications that were not in English, which might have caused the omission of some publications. However, considering that English is the lingua franca in the review of papers, this threat can be ignored.

### *B. Mistake Exclusion for Publications*

Despite our strict exclusion criteria for white and grey literature, there may still be cases where publications are excluded by mistake. To address this, we excluded literature through an incremental review, examining publications by title and abstract, sweeping the full text, and reading the full text. In addition, the authors of this SLR flagged publications individually, and informal discussions were used to obtain final results for divergent publications, which also helped to mitigate publication exclusions by mistake.

### *C. Reliability of this SLR*

We are using Fleiss Kappa method to help determine the extent of researcher agreement during the screening phase of this SRL [167]. After removing duplicates, we allocated 25 publications to three researchers randomly selected from the database to implement the Fleiss kappa analysis. The three researchers were required to speak about publications judged to be included, excluded, or not known. After the above process, we arrived at an average agreement of 84% between the researchers' judgments and the results of this SLR [167], which indicates a high degree of agreement between the researchers' judgments and the results of this SLR. This method mitigates possible reliability threats. In addition, we mitigate potential threats to the reliability of extracted data by systematically documenting it using well-defined processes and providing a database that can replicate each process step. The database is open source accessible[1].

## VII. Related Work

Schedulability analysis in TSN has consistently been a focal point of research in the field. However, few reviews have comprehensively summarized the relevant literature on this topic, resulting in a persistent lack of systematic review. A systematic review is essential for thoroughly analyzing the issues associated with TSN schedulability analysis.

Deng *et al.* [127] summarized articles on modeling from AVB to TSN, end-to-end delay analysis, real-time scheduling, reliability-aware design, security-aware design, TSN in the automotive domain, and provided insights into future trends. Regarding schedulability analysis, the paper distinguishes between the traffic shaping mechanisms of AVB and TSN, emphasizing the unique characteristics of methods used in various studies. However, a limitation of this review is that it includes summaries of only approximately 20 articles related to schedulability analysis.

Similarly, Ashjaei *et al.* [126], focusing on research in automotive time-sensitive networking, provide a comprehensive review. This review includes a more extensive summary of relevant articles in the schedulability analysis section and identifies future research directions. Additionally, the article explores the evolution of TSN in automotive embedded system applications, model-based software development of automotive distributed embedded systems utilizing TSN, simulation platforms for TSN, hardware evolution in TSN, and considerations of safety and security in TSN. Houtan et al. [98] conducted schedulability analysis of

[1] https://github.com/Zitong-W/Database-schedulability-analysis-in-TSN

BE traffic in TSN networks. In its related work section, this work summarized four methods for schedulability analysis, focusing on the NC approach, the Response Time Analysis (RTA) approach, the Machine Learning (ML) approach, and the EI approach. The work also included a timeline of schedulability analysis techniques for TSN since 2014. However, like [127], these two work did not comprehensively cover research related to schedulability analysis. In addition, the work in [126] did not categorize the studies according to the methods of schedulability analysis.

Maile *et al.* [168] introduced several TSN standards and elucidated the fundamental concepts and algorithms of the NC approach. It argues that the NC approach can establish reliable upper bounds for end-to-end delays in TSN, albeit potentially conservative. The paper also reviews current literature on NC, providing valuable insights for readers interested in the application of NC within TSN contexts. Unfortunately, this review only provides an overview of studies using the NC approach and covers a less comprehensive literature.

Moreover, Bello and Steiner [169] reviewed and analyzed the IEEE 802.1 TSN framework within the context of industrial communication and automation systems. They offered an overview of current projects and critically examine key TSN standards deemed crucial for industrial applications. Additionally, they explored application domains, with particular emphasis on industrial automation and automotive sectors. Given the emerging nature of TSN, academic research in this area is gaining momentum. However, since this review was published in 2019, its coverage of TSN standards is not comprehensive, for instance, it lacks an introduction to the ATS mechanism. Craciunas *et al.* [129] outlined the core scheduling principles aimed at achieving deterministic behavior for critical streams within Time-Triggered Ethernet (TTE) and TSN networks. This review introduced modeling guidelines for traffic and network structures, and explored diverse global and local constraint conditions. These constraints are leveraged for network scheduling through SMT. Furthermore, the review delves into potential criteria for optimization. However, these two works did not describe and summarize the schedulability analysis research in TSN.

In conclusion, although several current reviews address the work on TSN and schedulability analysis, there still needs to be a comprehensive overview of schedulability analysis in TSN and a structured research map of the field. Therefore, through SLR, we present all the current research conducted within the scope of schedulability analysis in TSN and identify further research opportunities for researchers.

## VII. Conclusion

This paper presents a systematic literature review for schedulability analysis in TSN and shows its planning scheme, execution process, and final results. Our search in various databases and retrieval systems yielded 786 publications. After our iterative filtering and elimination process, we narrowed the number to 123 publications, of which eight were obtained from the snowballing approach. These publications focus on schedulability analyses of traffic under various scheduling mechanisms in TSN. We extracted and analyzed data from these publications through clear definitions and processes to answer the four research questions posed in this SLR.

Before answering the four research questions, we conducted a preliminary analysis of these publications. The study results show that most of the research has been published by the IEEE at conferences, with the IEEE Conference on Emerging Technologies & Factory Automation (ETFA) publishing the most primary publication. The interest of researchers in this area is growing. Luxi Zhao has done the most research in this area, and many other researchers are gradually becoming involved.

To address the first question, we analyzed and classified the research objectives of the publications. Through the analysis results, we can see that all publications are analyzed for schedulability to provide tight boundaries. In addition, 45 publications (37% of the total) also aimed at evaluating the effectiveness of the mechanisms or comparing them with each other by using the results of the schedulability analysis. In contrast, 24 (20% of the total) and 17 (14% of the total) used the schedulability analysis to evaluate the feasibility of the configurations (or to obtain configuration recommendations) and to construct the constraints used in the solution, respectively.

To answer the second question, we categorized the publications based on explicit classification criteria of TSN scheduling mechanisms and traffic types. It is shown that current publications addressing schedulability analysis of flows under the CBS and TAS scheduling mechanisms are overwhelmingly dominant (their sum accounts for more than 80% of the total). Research on other scheduling mechanisms and the integration of scheduling mechanisms still needs to be expanded. At the same time, research on CBS and TAS has led to a correspondingly high number of publications analyzing CDT and AVB traffic. Similarly, schedulability analyses for BE traffic still need attention.

The third research question is to classify publications based on the schedulability analysis methodology and assess the extensiveness of their analyses. The results of the study show that the method used for schedulability analysis in most of the current publications is the NC approach (42% of the total), followed by the FTA approach (23% of the total) and the BPA approach (22% of the total). However, the inability of the FTA approach to provide a well-developed analytical framework led to a lower quality of modeling and analysis in some of the publications. In addition, we list the extended relationships between different publications to facilitate a more intuitive presentation of the optimization history of various schedulability analysis methods.

To answer the last question, we categorize the assessment methods and domains used in the publications. From the classification results, it can be seen that most of the current publications tend to combine the analysis with computational results or through simple examples (50% of the total), followed by the evaluation of the analysis results through simulation (37% of the total), with the most commonly used simulation platform being OMNeT++ (23 publications). In addition, most current publications use generic scenarios in their evaluation (57 publications), with at most 30% of the total using network scenarios from the automotive, industrial, and aerospace domains.

After answering the research questions, we further analyzed the trends of publications under various classifications, obtaining information on the evolution of the number of publications for various analysis purposes, scheduling mechanisms, traffic types, analysis methods, and evaluation methods. In addition, we explored the connection between publications under different classifications through bubble charts. The analyses suggest that research targeting mechanisms or combinations of mechanisms other than CBS and TAS through different scheduling methods is also a direction for future research work. We also show the relationship between different scheduling methods and the extensiveness of analysis, and the results demonstrate the inability of FTA to provide a complete analytical framework. We also show the relationship between the expansion and comparison of different analysis methods to demonstrate the connection between analysis methods and to show that there is still room for optimization of various current analysis methods. Finally, we list open issues and future work to point out current directions in the field that remain to be investigated.

## APPENDIX

A complete list of all publications is given in Table A1.

TABLE A1
FULL LIST OF THE PUBLICATIONS

| No. | Cit. | Title | Year | Type |
|---|---|---|---|---|
| 1 | [35] | Adaptive Priority Adjustment Scheduling Approach with Response-Time Analysis in Time-Sensitive Networks | 2022 | Journal |
| 2 | [96] | Formal timing analysis of non-scheduled traffic in automotive scheduled TSN networks | 2017 | Conference |
| 3 | [36] | Formal worst-case timing analysis of ethernet TSN's burst-limiting shaper | 2016 | Conference |
| 4 | [37] | Incorporating TSN/BLS in AFDX for mixed-criticality applications: Model and timing analysis | 2018 | Conference |
| 5 | [38] | Worst-Case Timing Analysis of AFDX Networks with Multiple TSN/BLS Shapers | 2020 | Journal |
| 6 | [39] | Formal worst-case timing analysis of Ethernet TSN's time-aware and peristaltic shapers | 2015 | Conference |
| 7 | [1] | Queue assignment for fixed-priority real-time flows in time-sensitive networks: Hardness and algorithm | 2021 | Journal |
| 8 | [40] | Schedulability analysis of Time-Sensitive Networks with scheduled traffic and preemption support | 2020 | Journal |
| 9 | [33] | Tight worst-case response-time analysis for ethernet AVB using eligible intervals | 2016 | Conference |
| 10 | [114] | Deadline-Aware Online Scheduling of TSN Flows for Automotive Applications | 2023 | Journal |
| 11 | [41] | Latency Analysis of Multiple Classes of AVB Traffic in TSN With Standard Credit Behavior Using Network Calculus | 2021 | Journal |
| 12 | [97] | Worst-Case Delay Bounds in Time-Sensitive Networks with Packet Replication and Elimination | 2022 | Journal |
| 13 | [73] | Independent WCRT analysis for the Best-Effort class BE in Ethernet AVB | 2022 | Conference |
| 14 | [42] | Fixed-priority scheduling and controller co-design for time-sensitive networks | 2020 | Conference |
| 15 | [74] | Schedulability analysis of Ethernet AVB switches | 2014 | Conference |
| 16 | [43] | Quantitative Performance Comparison of Various Traffic Shapers in Time-Sensitive Networking | 2022 | Journal |
| 17 | [2] | Hierarchical scheduling and real-time analysis for vehicular time-sensitive network | 2019 | Conference |
| 18 | [44] | Formal worst-case performance analysis of time-sensitive Ethernet with frame preemption | 2016 | Conference |
| 19 | [75] | Independent yet Tight WCRT Analysis for Individual Priority Classes in Ethernet AVB | 2016 | Conference |
| 20 | [45] | Schedulability analysis of Ethernet Audio Video Bridging networks with scheduled traffic support | 2017 | Journal |
| 21 | [120] | Timing Analysis of AVB Traffic in TSN Networks Using Network Calculus | 2018 | Conference |
| 22 | [76] | Independent WCRT analysis for individual priority classes in Ethernet AVB | 2018 | Journal |
| 23 | [77] | Timing analysis of AVB Ethernet network using the Forward end-to-end Delay Analysis | 2018 | Conference |
| 24 | [115] | Worst-case traversal time analysis of TSN with multi-level preemption | 2021 | Journal |
| 25 | [46] | Network Calculus-based Timing Analysis of AFDX networks with Strict Priority and TSN/BLS Shapers | 2018 | Conference |
| 26 | [47] | Worst-Case Latency Analysis for AVB Traffic Under Overlapping-Based Time-Triggered Windows in Time-Sensitive Networks | 2022 | Journal |
| 27 | [48] | Worst-Case Latency Analysis for IEEE 802.1Qbv Time Sensitive Networks Using Network Calculus | 2018 | Journal |
| 28 | [98] | Schedulability Analysis of Best-Effort Traffic in TSN Networks | 2021 | Conference |
| 29 | [49] | Analysis of TSN for Industrial Automation based on Network Calculus | 2019 | Conference |
| 30 | [99] | Run-time Per-Class Routing of AVB Flows in In-Vehicle TSN via Composable Delay Analysis | 2022 | Conference |
| 31 | [50] | Network calculus-based latency for time-triggered traffic under flexible window-overlapping scheduling (Fwos) in a time-sensitive network (tsn) | 2021 | Journal |
| 32 | [3] | Comparison of Time Sensitive Networking (TSN) and TTEthernet | 2018 | Conference |
| 33 | [51] | Critical Offset Optimizations for Overlapping-Based Time-Triggered Windows in Time-Sensitive Network | 2021 | Journal |
| 34 | [78] | Deterministic delay analysis of AVB switched Ethernet networks using an extended Trajectory Approach | 2017 | Journal |
| 35 | [15] | Improving worst-case delay analysis for traffic of additional stream reservation class in ethernet-AVB network | 2018 | Journal |
| 36 | [16] | Analysis of ethernet-switch traffic shapers for in-vehicle networking applications | 2015 | Conference |
| 37 | [17] | IEC 61850 over TSN: traffic mapping and delay analysis of GOOSE traffic | 2020 | Conference |
| 38 | [52] | Delay Analysis of TSN Based Industrial Networks with Preemptive Traffic Using Network Calculus | 2023 | Conference |
| 39 | [79] | Comparison of AFDX and audio video bridging forwarding methods using network calculus approach | 2017 | Conference |
| 40 | [53] | Improving Worst-case TSN Communication Times of Large Sensor Data Samples by Exploiting Synchronization | 2023 | Journal |
| 41 | [4] | Worst-Case Response Time Analysis for Best- Effort Traffic in an Ethernet-AVB Network | 2021 | Conference |
| 42 | [18] | Timing Analysis for Optimal Points in Credit-Based Shaper of Time Sensitive Network | 2021 | Conference |
| 43 | [107] | Research on Cyclic Queuing and Forwarding with Preemption in Time-Sensitive Networking | 2024 | Journal |
| 44 | [80] | Timing analysis of Ethernet AVB-based automotive E/E architectures | 2013 | Conference |
| 45 | [54] | Formal worst-case timing analysis of Ethernet topologies with strict-priority and AVB switching | 2012 | Conference |
| 46 | [55] | Network Calculus-based Timing Analysis of AFDX networks incorporating multiple TSN/BLS traffic classes | 2019 | Journal |
| 47 | [19] | Formal and simulation-based timing analysis of industrial-ethernet sercos III over TSN | 2017 | Conference |

| 48 | [108] | A Traffic Scheduling Algorithm Combined with Ingress Shaping in TSN | 2022 | Conference |
|---|---|---|---|---|
| 49 | [56] | Improving Latency Analysis for Flexible Window-Based GCL Scheduling in TSN Networks by Integration of Consecutive Nodes Offsets | 2021 | Journal |
| 50 | [109] | Delay analysis of AVB traffic in time-sensitive networks (TSN) | 2017 | Conference |
| 51 | [20] | Data Flow Control for Network Load Balancing in IEEE Time Sensitive Networks for Automation | 2023 | Journal |
| 52 | [81] | On the use of supervised machine learning for assessing schedulability: application to ethernet TSN | 2019 | Conference |
| 53 | [100] | Bandwidth Allocation of Stream-Reservation Traffic in TSN | 2022 | Journal |
| 54 | [82] | SynAVB: Route and Slope Synthesis Ensuring Guaranteed Service in Ethernet AVB | 2022 | Conference |
| 55 | [83] | Exploiting Shaper Context to Improve Performance Bounds of Ethernet AVB Networks | 2014 | Conference |
| 56 | [101] | Improving formal timing analysis of switched ethernet by exploiting FIFO scheduling | 2015 | Conference |
| 57 | [121] | An Analytical Latency Model for AVB Traffic in TSN Considering Time-Triggered Traffic | 2020 | Conference |
| 58 | [102] | AVB-aware Routing and Scheduling for Critical Traffic in Time-sensitive Networks with Preemption | 2022 | Conference |
| 59 | [57] | Real-time guarantees for critical traffic in ieee 802.1 qbv tsn networks with unscheduled and unsynchronized end-systems | 2021 | Journal |
| 60 | [5] | Deep learning to predict the feasibility of priority-based ethernet network configurations | 2021 | Journal |
| 61 | [84] | A Hybrid Machine Learning and Schedulability Analysis Method for the Verification of TSN Networks | 2019 | Conference |
| 62 | [119] | Improvements to Deep-Learning-based Feasibility Prediction of Switched Ethernet Network Configurations | 2021 | Conference |
| 63 | [21] | Routing optimization of AVB streams in TSN networks | 2016 | Journal |
| 64 | [58] | Urgency-Based Scheduler for Time-Sensitive Switched Ethernet Networks | 2016 | Conference |
| 65 | [85] | Complete modelling of AVB in Network Calculus Framework | 2014 | Conference |
| 66 | [86] | Analysis of Ethernet AVB for automotive networks using Network Calculus | 2012 | Conference |
| 67 | [87] | An independent yet efficient analysis of bandwidth reservation for credit-based shaping | 2018 | Conference |
| 68 | [59] | Incorporating TSN/BLS in AFDX for mixed-criticality avionics applications: Specification and analysis | 2017 | Journal |
| 69 | [88] | Latency and Backlog Bounds in Time-Sensitive Networking with Credit Based Shapers and Asynchronous Traffic Shaping | 2018 | Conference |
| 70 | [89] | Impact analysis of flow shaping in ethernet-AVB/TSN and AFDX from network calculus and simulation perspective | 2017 | Journal |
| 71 | [117] | AVB-Aware Routing and Scheduling of Time-Triggered Traffic for TSN | 2018 | Journal |
| 72 | [60] | Synthesis of Queue and Priority Assignment for Asynchronous Traffic Shaping in Switched Ethernet | 2017 | Conference |
| 73 | [61] | A Feasibility Analysis Framework of Time-Sensitive Networking Using Real-Time Calculus | 2019 | Journal |
| 74 | [22] | Research on Automotive TSN network Scheduling Analysis and Simulation | 2023 | Conference |
| 75 | [6] | Analysis and modeling of asynchronous traffic shaping in time sensitive networks | 2018 | Conference |
| 76 | [7] | Analyzing and modeling the latency and jitter behavior of mixed industrial TSN and DetNet networks | 2022 | Conference |
| 77 | [103] | Bandwidth Reservation Analysis for Schedulability of AVB Traffic in TSN | 2024 | Conference |
| 78 | [116] | Bounding the Upper Delays of the Tactile Internet Using Deterministic Network Calculus | 2023 | Journal |
| 79 | [23] | The Converged Scheduling for Time Sensitive Mission in Satellite Formation Flying | 2022 | Conference |
| 80 | [24] | Delay Analysis and Testing of TAS Mechanism in Time-Sensitive Networks | 2023 | Conference |
| 81 | [104] | The Delay Bound Analysis Based on Network Calculus for Asynchronous Traffic Shaping under Parameter Inconsistency | 2020 | Conference |
| 82 | [25] | Delay optimization strategy based on aperiodic traffic in time-sensitive networking | 2021 | Conference |
| 83 | [90] | Delay-Guaranteeing Admission Control for Time-Sensitive Networking Using the Credit-Based Shaper | 2022 | Journal |
| 84 | [26] | Design and implementation of a frame preemption model without guard bands for time-sensitive networking | 2024 | Journal |
| 85 | [27] | A Design of Token Bucket Shaper Aided with Gate Control List in Time-Sensitive Networks | 2022 | Conference |
| 86 | [62] | Differential Traffic QoS Scheduling for 5G/6G Fronthaul Networks | 2021 | Conference |
| 87 | [8] | Egress-TT Configurations for TSN Networks | 2022 | Conference |
| 88 | [110] | End-to-End Delay Analysis of ATS Mechanism at Edge Nodes Via Network Calculus | 2023 | Conference |
| 89 | [28] | Enhanced Frame Preemption in Image and Video Transmission Over Time Sensitive Networks | 2022 | Conference |
| 90 | [111] | Enhanced Real-time Scheduling of AVB Flows in Time-Sensitive Networking | 2024 | Journal |
| 91 | [63] | Improve Service Curve Using Non-overlapped Gate in Time Sensitive Network Switch | 2021 | Conference |
| 92 | [9] | Joint Algorithm of Message Fragmentation and No-Wait Scheduling for Time-Sensitive Networks | 2021 | Journal |
| 93 | [105] | MALOC: Building an adaptive scheduling and routing framework for rate-constrained TSN traffic | 2022 | Conference |
| 94 | [91] | Minimum Bandwidth Reservation for CBS in TSN With Real-Time QoS Guarantees | 2024 | Journal |
| 95 | [64] | Modeling and Analysis of Time-Aware Shaper on Half-Duplex Ethernet PLCA Multidrop | 2023 | Journal |
| 96 | [65] | Network Calculus-based Modeling of Time Sensitive Networking Shapers for Industrial Automation Networks | 2019 | Conference |
| 97 | [10] | On the performance of stream-based class-based time-aware shaping and frame preemption in TSN | 2020 | Conference |
| 98 | [92] | On the Validity of Credit-Based Shaper Delay Guarantees in Decentralized Reservation Protocols | 2023 | Conference |
| 99 | [29] | Online Rerouting and Rescheduling of Time-Triggered Flows for Fault Tolerance in Time-Sensitive Networking | 2022 | Journal |
| 100 | [66] | Performance enhancement of extended AFDX via bandwidth reservation for TSN/BLS shapers | 2019 | Journal |
| 101 | [67] | Priority Re-assignment for Improving Schedulability and Mixed-Criticality of ARINC 664 | 2021 | Conference |
| 102 | [11] | Real-Time Scheduling of Massive Data in Time Sensitive Networks with a Limited Number of Schedule Entries | 2020 | Journal |
| 103 | [68] | Research on Maximum Delay of Time-sensitive Network Based on GCL Adaptive Adjustment | 2023 | Conference |
| 104 | [12] | reTSN: Resilient and Efficient Time-Sensitive Network for Automotive In-Vehicle Communication | 2023 | Journal |
| 105 | [30] | Robust Time-Sensitive Networking with Delay Bound Analyses | 2021 | Conference |
| 106 | [106] | A SDN-based Traffic Bandwidth Allocation Method for Time Sensitive Networking in Avionics | 2019 | Conference |
| 107 | [112] | Some Basic Properties of Length Rate Quotient | 2023 | Conference |
| 108 | [13] | Synchronous Time-Sensitive Networking Scheduling Algorithm Based on Dynamic Time Margin | 2023 | Conference |
| 109 | [113] | Token Regulated Credit Based Shaper for Time Sensitive Networks | 2020 | Conference |
| 110 | [118] | A Topology-specific Tight Worst-case Analysis of Strict Priority Traffic in Real-time Systems | 2023 | Conference |
| 111 | [69] | Upper Bound Analysis of TSN End-to-End Delay Based on Network Calculus | 2023 | Conference |
| 112 | [70] | uTAS: Ultra-Reliable Time-Aware Shaper for Time-Sensitive Networks | 2023 | Conference |
| 113 | [71] | Window based Dynamic Scheduling Algorithm in Time Sensitive Networks | 2022 | Conference |
| 114 | [31] | Worst-case analysis of ethernet AVB in automotive system | 2015 | Conference |

| 115 | [32] | Worst-Case Delay Slicing for Time-Sensitive Applications in Softwarized Industrial Networks | 2020 | Conference |
|---|---|---|---|---|
| 116 | [93] | Modeling of Ethernet AVB networks for worst-case timing analysis | 2012 | Journal |
| 117 | [122] | Modelling in network calculus a TSN architecture mixing Time-Triggered Credit Based Shaper and Best-Effort queues | 2018 | Journal |
| 118 | [123] | Improving worst-case end-to-end delay analysis of multiple classes of AVB traffic in TSN networks using network calculus | 2018 | Journal |
| 119 | [34] | Towards the calculation of performance guarantees for BLS in Time-Sensitive Networks | 2013 | Conference |
| 120 | [94] | Improved credit bounds for the credit-based shaper in time-sensitive networking | 2019 | Journal |
| 121 | [95] | Using machine learning to speed up the design space exploration of Ethernet TSN networks | 2019 | Journal |
| 122 | [72] | Specification and analysis of an extended AFDX with TSN/BLS shapers for mixed-criticality avionics applications | 2018 | Thesis |
| 123 | [14] | Bounded latency with bridge-local stream reservation and strict priority queuing | 2020 | Conference |